\documentclass[a4paper,11pt]{article}
\pdfoutput=1 

\usepackage{jheppub} 

\usepackage{hyperref}
\usepackage[utf8]{inputenc}
\usepackage[T1]{fontenc}
\usepackage{amsmath}
\usepackage{amssymb}
\usepackage{wrapfig}
\usepackage{feynmp}
\usepackage[english]{babel} 
\usepackage{array}


\makeatletter
\let\save@mathaccent\mathaccent
\newcommand*\if@single[3]{%
  \setbox0\hbox{${\mathaccent"0362{#1}}^H$}%
  \setbox2\hbox{${\mathaccent"0362{\kern0pt#1}}^H$}%
  \ifdim\ht0=\ht2 #3\else #2\fi
  }
\newcommand*\rel@kern[1]{\kern#1\dimexpr\macc@kerna}
\newcommand*\widebar[1]{\@ifnextchar^{{\wide@bar{#1}{0}}}{\wide@bar{#1}{1}}}
\newcommand*\wide@bar[2]{\if@single{#1}{\wide@bar@{#1}{#2}{1}}{\wide@bar@{#1}{#2}{2}}}
\newcommand*\wide@bar@[3]{%
  \begingroup
  \def\mathaccent##1##2{%
    \let\mathaccent\save@mathaccent
    \if#32 \let\macc@nucleus\first@char \fi
    \setbox\z@\hbox{$\macc@style{\macc@nucleus}_{}$}%
    \setbox\tw@\hbox{$\macc@style{\macc@nucleus}{}_{}$}%
    \dimen@\wd\tw@
    \advance\dimen@-\wd\z@
    \divide\dimen@ 3
    \@tempdima\wd\tw@
    \advance\@tempdima-\scriptspace
    \divide\@tempdima 10
    \advance\dimen@-\@tempdima
    \ifdim\dimen@>\z@ \dimen@0pt\fi
    \rel@kern{0.6}\kern-\dimen@
    \if#31
      \overline{\rel@kern{-0.6}\kern\dimen@\macc@nucleus\rel@kern{0.4}\kern\dimen@}%
      \advance\dimen@0.4\dimexpr\macc@kerna
      \let\final@kern#2%
      \ifdim\dimen@<\z@ \let\final@kern1\fi
      \if\final@kern1 \kern-\dimen@\fi
    \else
      \overline{\rel@kern{-0.6}\kern\dimen@#1}%
    \fi
  }%
  \macc@depth\@ne
  \let\math@bgroup\@empty \let\math@egroup\macc@set@skewchar
  \mathsurround\z@ \frozen@everymath{\mathgroup\macc@group\relax}%
  \macc@set@skewchar\relax
  \let\mathaccentV\macc@nested@a
  \if#31
    \macc@nested@a\relax111{#1}%
  \else
    \def\gobble@till@marker##1\endmarker{}%
    \futurelet\first@char\gobble@till@marker#1\endmarker
    \ifcat\noexpand\first@char A\else
      \def\first@char{}%
    \fi
    \macc@nested@a\relax111{\first@char}%
  \fi
  \endgroup
}

\makeatletter \@addtoreset{equation}{section} \@addtoreset{equation}{subsection} \makeatother
\numberwithin{equation}{section}

\newcolumntype{M}[1]{>{\centering\arraybackslash}m{#1}}
\newcolumntype{N}{@{}m{0pt}@{}}

\newcommand{\ba}{\begin{array}}
\newcommand{\ea}{\end{array}}
\newcommand{\be}{\begin{equation}}
\newcommand{\ee}{\end{equation}}
\newcommand{\bea}{\begin{eqnarray}}
\newcommand{\eea}{\end{eqnarray}}
\newcommand{\bg}{\begin{gather}}
\newcommand{\eg}{\end{gather}}
\newcommand{\bseq}{\begin{subequations}}
\newcommand{\eseq}{\end{subequations}}

\makeatletter

\def\lsim{\compoundrel<\over\sim}
\def\compoundrel#1\over#2{\mathpalette\compoundreL{{#1}\over{#2}}}
\def\compoundreL#1#2{\compoundREL#1#2}
\def\compoundREL#1#2\over#3{\mathrel
         {\vcenter{\hbox{$\m@th\buildrel{#1#2}\over{#1#3}$}}}}
\makeatother

\preprint{INR-TH-2016-015}
\title{Lepton flavor-violating decays of the Higgs boson \\ from sgoldstino mixing}
\author[a]{S.V. Demidov,}
\author[a,b]{I.V. Sobolev}

\affiliation[a]{Institute for Nuclear Research of the Russian Academy of Sciences,\\ 60th October Anniversary prospect 7a, Moscow 117312, Russia}
\affiliation[b]{Department of Particle Physics and Cosmology, Physics Faculty, \\ M.~V.~Lomonosov Moscow State University,\\ Vorobjevy Gory, 119991, Moscow, Russia}

\emailAdd{demidov@ms2.inr.ac.ru}
\emailAdd{sobolev.ivan@physics.msu.ru}

\abstract{ We study lepton flavor violation in a class of supersymmetric models 
  with light sgoldstino -- scalar superpartner of Goldstone fermions
  responsible for spontaneous supersymmetry breaking. Sgoldstino
  couplings to the Standard Model (SM) fermions are determined by the
  MSSM soft terms and, in general, provide with flavor
  violation in this sector. Sgoldstino admixture to the lightest Higgs
  boson results in changes of its coupling constants and, in particular,
  leads to lepton flavor-violating decay $h\to\tau\mu$ of the Higgs
  resonance. We discuss viability and phenomenological consequences
  of this scenario. }

\keywords{Supersymmetry Phenomenology}

\arxivnumber{1605.08220}

\begin{document}
\maketitle
\toccontinuoustrue


\section{Introduction}
Study of the Higgs boson properties is one of the priority problems
since its discovery~\cite{Aad:2012tfa,Chatrchyan:2012xdj}. Special
attention is paid to flavor changing processes involving the Higgs
boson and, in particular, to the lepton flavor-violating (LFV) Higgs
boson decays~\cite{Khachatryan:2015kon,Aad:2015gha}. Prospects of
studying the LFV Higgs boson decays at the LHC experiments were
discussed in~\cite{Blankenburg:2012ex,Harnik:2012pb,
Kopp:2014rva,DiazCruz:2008ry,Pilaftsis:1992st}. In 
particular, it has been found that given strong constraints from FCNC
physics sufficiently large branching ratios of the Higgs boson decays $h\to
e\tau$ and $h\to\mu\tau$ are still allowed\footnote{Here we denote
  ${\rm Br}(h\to l_il_j) \equiv {\rm Br}(h\to l_i\widebar{l_j}) + {\rm 
    Br}(h\to\widebar{l_i}l_j)$}. Recently, the latter
$h\to\mu\tau$ decay has drawn much attention because the latest
results for upper limits on the branching ratio of $h\to\tau\mu$ decay
have been reported by the ATLAS, ${\rm Br}(h\to\tau\mu)< 1.85\times
10^{-2}$, and CMS, ${\rm Br}(h\to\tau\mu) < 1.51\times 10^{-2}$ at
95\% CL. At the same time, the CMS analysis revealed a small excess in
this process with a significance of 2.4$\sigma$ which can be
interpreted as LFV Higgs decay with branching ${\rm
  Br}(h\to\tau\mu)=8.4^{+3.9}_{-3.7}\times 10^{-3}$. Although not yet
statistically significant, this excess is very intriguing. If
confirmed, it would give direct indication on non-SM properties of the
Higgs boson. 

\noindent
To explain this excess, various models of new physics have been
studied,
including~\cite{Dery:2014kxa,Campos:2014zaa,Sierra:2014nqa,Heeck:2014qea,Crivellin:2015mga,Dorsner:2015mja,He:2015rqa,Altmannshofer:2015esa,Arganda:2015naa,Botella:2015hoa,Baek:2015mea,Buschmann:2016uzg}. In
what follows, we will be interested in supersymmetric scenarios. LFV
decays of the Higgs boson in MSSM was discussed
\footnote{See Refs.~\cite{Arganda:2004bz,DiazCruz:2002er,DiazCruz:1999xe} for studies of LFV  Higgs boson decays in other supersymmetric models.}
in~\cite{Brignole:2003iv,Brignole:2004ah} and recently
in~\cite{Arana-Catania:2013xma}. Previous studies of $h\to\mu\tau$
decay with account of the CMS excess can be found in
Refs.~\cite{Arganda:2015uca},~\cite{Aloni:2015wvn}.  These studies
revealed that for a generic set of parameters predictions
for ${\rm Br}(h\to\mu\tau)$ are very small for this decay to be
observed at the LHC and only limited parameter space of MSSM is
capable of explaining the CMS excess. 

\noindent
In this paper, we will be interested in explanation of the CMS excess
within the framework of a particular class of supersymmetric models with low scale supersymmetry breaking (see, e.g~\cite{Brignole:2003cm}
and recent studies in~\cite{Petersson:2011in,Antoniadis:2012ck,
Dudas:2012fa,Dudas:2013mia,Das:2015zwa}. 
In these models, it is assumed that the scale of supersymmetry
breaking $\sqrt{F}$ is not very far from the electroweak energy scale. In
this case, particles responsible for the spontaneous supersymmetry
breaking may show up already in the LHC
experiments~\cite{Shirai:2009kn,Klasen:2006kb,Brignole:1998me,Petersson:2012dp,Gorbunov:2000ht,Gorbunov:2002er,Demidov:2004qt,Petersson:2015rza}. This
additional sector contains goldstino and its superpartners --
sgoldstinos, which in the simplest case are scalars. The coupling
constants of this sector to the SM particles are governed by the soft 
supersymmetry breaking parameters of supersymmetric model which are 
generally flavor violating and can lead to FCNC
processes~\cite{Gorbunov:2000th,Brignole:2000wd,Gorbunov:2000cz,Gorbunov:2005nu,Demidov:2006pt,Demidov:2011rd,Astapov:2015otc}. The
main idea for the explanation of the CMS excess is that the Higgs
boson can mix with the scalar 
sgoldstino~\cite{Bellazzini:2012mh,Astapov:2014mea}, while the latter
has flavor-violating interactions with the SM fermions. 

\noindent
Below we calculate the contribution of the sgoldstino-Higgs mixing to 
$h\to\tau\mu$ decay and analyze constraints from relevant FCNC
processes and from LHC data. We find that this mixing is capable of
explaining the CMS excess in a part of the parameter space. Also we
discuss possible implications of this scenario for the Higgs boson
physics as well as for several FCNC processes. In Section~2 we describe the
theoretical framework of low scale supersymmetry breaking models and
discuss sgoldstino-Higgs mixing. In Section~3 we turn to the
phenomenological analysis, performing a scan over relevant parameter
space of the model and discussing experimental constraints. In
Section~4 we present the results of the scan and reveal interesting
features which can be useful to verifying this scenario. Section~5
contains our conclusions and several technical aspects are left for
appendices.

\section{Theoretical framework}

Here we briefly describe the main features of the supersymmetric model
with light sgoldstinos. In addition to the SM fields and their
superpartners of the conventional MSSM we introduce goldstino chiral
superfield $\Phi = \phi + \sqrt{2} \theta \widetilde{G} + F_{\phi}
\theta^2$. Here $\tilde{G}$ is the Goldstone fermion, $\phi$ is the
sgoldstino field and $F_\phi$ is the auxiliary field. Due to some
dynamics in the hidden sector, the field $F_{\phi}$ acquires vacuum
expectation value which breaks SUSY spontaneously. We restrict
ourselves to the simplest set of operators which reproduces soft
SUSY-breaking parameters of MSSM after spontaneous supersymmetry
breaking~\cite{Petersson:2011in,Gorbunov:2001pd}. We use the following
lagrangian 
\begin{equation}
\label{Lagr1}
\mathcal{L}_{\text{model}} = \mathcal{L}_{\text{K{\"a}hler}} +
\mathcal{L}_{\text{superpotential}}.
\end{equation}
The contribution to the K{\"a}hler potential has the form  
\begin{equation}
\label{Kahler}
\mathcal{L}_{\text{K{\"a}hler}} = \int \, d^2 \theta \, d^2
\widebar{\theta} \, \sum_{k} \left (1- \frac{m_k^2}{F^2}\Phi^{\dagger}
\Phi \right) \Phi^{\dagger}_k e^{g_1 V_1+g_2 V_2+g_3 V_3} \Phi_k ,
\end{equation}
where the sum goes over all chiral MSSM superfields and we implicitly
assume possibility of nontrivial flavor structure for the soft
parameters $m_{k}^2$ of sleptons and squarks. The contribution from
the superpotential is  
\begin{eqnarray}
\label{Superpotential}
\begin{aligned}
\mathcal{L}_{\text{superpotential}} = & \int \, d^2 \theta \, \Biggl
\{ \epsilon_{ij} \Biggl( \left( \mu - \frac{B}{F}\Phi\right) H_{d}^{i}
H_{u}^{j} + \left(Y_{ab}^{L}+\frac{A_{ab}^{L}}{F} \Phi \right)
L_{a}^{j} E_{b}^{c} H_{d}^{i} + \\ 
& + \left(Y_{ab}^{D}+\frac{A_{ab}^{D}}{F} \Phi \right) Q_{a}^{j}
D_{b}^{c} H_{d}^{i} + \left(Y_{ab}^{U}+\frac{A_{ab}^{U}}{F} \Phi
\right) Q_{a}^{i} U_{b}^{c} H_{d}^{j} \Biggr) + \\ 
& + \frac{1}{4} \sum_{\alpha} \left(1+\frac{2M_{\alpha}}{F}\Phi
\right) \text{Tr} \, W^{\alpha} W^{\alpha} \Biggr \} + \text{h.c.} 
\end{aligned}
\end{eqnarray}
Here $B$, $A_{ab}^{L,D,U}$ and $M_{\alpha}$, $a,b,\alpha=1,2,3$ are the
soft MSSM parameters. For lagrangian of the hidden sector, we choose
the following form 
\begin{equation}
\label{SuperpF}
\mathcal{L}_{\Phi} = \int \, d^2 \theta \, d^2 \widebar{\theta} \left(
\Phi^{\dagger}\Phi + \widetilde{K} (\Phi^{\dagger},\Phi)\right) -
\left( \int d^2 \theta F \Phi + \text{h.c.} \right), 
\end{equation}
where the first term is the canonical kinetic term while the second
one, $\widetilde{K} (\Phi^{\dagger},\Phi)$, represents some
complicated dynamics in the hidden sector and is suppressed by powers of  
$F$. The last linear term in the superpotential of Eq.~(\ref{SuperpF}) 
forces the auxiliary field $F_{\phi}$ to acquire non-zero vacuum expectation
value $\langle F_{\phi} \rangle = F + \mathcal{O}
\left(\frac{1}{F}\right)$ and hence triggers spontaneous supersymmetry breaking. In what follows, we assume that all the parameters of the lagrangian (\ref{Lagr1}) -- (\ref{SuperpF}) are real and hence ignore possible CP-violation~\footnote{CP-violation in the Higgs boson decays has been discussed in Refs.~\cite{Kopp:2014rva},~\cite{Korner:1992zk} in view of the experiments at the LHC. At the same time, complex flavour-violating Yukawa couplings can lead to non-zero electric dipole moment of muon~\cite{Harnik:2012pb}. We leave discussion of implications of these interesting effects in the framework of the model in question for future studies.}  
\noindent
After integrating out the auxiliary fields of sgoldstino and Higgs
chiral superfields as well as auxiliary fields of vector superfields
containing the SM gauge bosons and assuming that $\sqrt{F}$ is the
largest energy scale 
of the model, the potential of the Higgs sector can be written as an
expansion in powers of $1/F$ as follows
\begin{equation}
\label{modelV}
V_{\text{model}} = V_{\text{MSSM}} + V^{(1)} + V^{(2)} + ...\;,
\end{equation}
where $V_{\text{MSSM}}$ is the MSSM scalar potential \cite{Martin:1997ns}
\begin{eqnarray}
\label{MSSMV}
\begin{aligned}
V_{\text{MSSM}} = & \left(\vert \mu \vert^2 + m_{H_u}^2 \right) \left(
\vert H_{u}^{0} \vert^2 + \vert H_{u}^{+} \vert^2 \right) +
\left(\vert \mu \vert^2 + m_{H_d}^2 \right) \left( \vert H_{d}^{0}
\vert^2 + \vert H_{d}^{+} \vert^2 \right) + \\ 
& + \left (B \left( H_{u}^{+} H_{d}^{-} - H_{u}^{0} H_{d}^{0} \right)
+ \text{c.c.} \right) + \\ 
& + \frac{g_1^2+g_2^2}{8} \left (\vert H_{u}^{0} \vert^2 + \vert
H_{u}^{+} \vert^2 - \vert H_{d}^{0} \vert^2 - \vert H_{d}^{+} \vert^2
\right)^2 + \frac{g_1^2}{2} \left \vert H_{u}^{+} H_{d}^{0*}+H_{u}^{0}
H_{d}^{-*} \right \vert^2 .
\end{aligned}
\end{eqnarray}
$V^{(1)}$ contains part of the potential responsible for 
Higgs-sgoldstino mixing 
\begin{eqnarray}
\label{mixingV}
\begin{aligned}
V_{\text{mixing}} = & \frac{\phi}{F} \Bigl( \mu \, \left( m_{H_u}^2 +
m_{H_d}^2 \right) \left(H_{u}^{0} H_{d}^{0}\right)^* - \frac{g_1^2
  M_1+g_2^2 M_2}{8} \left( \vert H_{u}^{0} \vert^2-\vert H_{d}^{0}
\vert^2 \right)^2 -  \\ 
& - B \mu \left( \vert H_{u}^{0} \vert^2 + \vert H_{d}^{0} \vert^2
\right) \Bigr) + \text{h.c.} 
\end{aligned} 
\end{eqnarray}
The part $V^{(2)}$ contains, in particular, $1/F^2$ corrections to the
MSSM Higgs potential
\begin{equation}
\label{correction}
V^{(2)} = \frac{1}{F^2}\left|m_{H_u}^2H_u^{\dagger}H_u +
m_{H_d}^2H_d^{\dagger}H_d - B\epsilon_{ij}H_{u}^iH_d^j\right|^2 + ...
\end{equation}
Other contributions to the scalar potential in Eqs.~\eqref{modelV} and~\eqref{correction} include sgoldstino potential, contributions of higher orders in $1/F$ and nonlinear interactions with sgoldstino which are not relevant for the present analysis. 

\noindent
Next, we expand Higgs $H_{u,d}^{0}$ and sgoldstino $\phi$ fields
around their minima as follows   
\begin{eqnarray}
\label{fields}
\begin{aligned}
& H_{u}^{0} = v_{u} + \frac{1}{\sqrt{2}} \left(h \, \cos \alpha + H \,
  \sin \alpha \right) + \frac{\dot{\imath}}{\sqrt{2}} A \, \cos \beta ,
  \\  
& H_{d}^{0} = v_{d} + \frac{1}{\sqrt{2}} \left(-h \, \sin \alpha + H
  \, \cos \alpha \right) + \frac{\dot{\imath}}{\sqrt{2}} A \, \sin
  \beta, \\  
& \phi = \frac{1}{\sqrt{2}} \left(s + \dot{\imath} p \right),
\end{aligned}
\end{eqnarray}
where $v_{u,d} = \langle H_{u,d}^0 \rangle$ and $v \equiv
  \sqrt{v_u^2+v_d^2} = 174~\rm{GeV}$, $\tan \beta =
  \frac{v_u}{v_d}$. Mixing angle between gauge $({\rm Re}
  H_{u}^{0},\,{\rm Re} H_{d}^{0})$ and mass eigenstates $(h,\,H)$ is
  denoted by $\alpha$. By   convention, $h$ is assumed to be lighter
  than $H$. $A$ is CP-odd  neutral Higgs field, while $s$ and $p$ are
  scalar and pseudoscalar sgoldstino components. In what follows, we
will work in the 
decoupling limit, i.e. $m_A \gg m_h$, or, equivalently $\cos \, \alpha
\approx \sin \, \beta$, $\sin \, 
\alpha \approx - \cos \, \beta$.
Substituting (\ref{fields}) into (\ref{mixingV}) and holding only
quadratic terms, one gets the following mass matrix in scalar sector 
\begin{equation}
\label{massmatrix}
\mathcal{M}_s^2 = 
\left(\begin{matrix}
m_H^2 & 0 & \frac{Y}{F} \\ 0 & m_h^2 & \frac{X}{F} \\ 
\frac{Y}{F} & \frac{X}{F} & m_s^2 
\end{matrix} \right),
\end{equation}
where the off-diagonal terms are
\begin{eqnarray}
\begin{aligned}
& X = 2 \mu^3 \, v \, \sin \, 2 \beta + \frac{v^3}{2} (g_1^2 \, M_1 +
  g_2^2 \, M_2) \cos^2 \, 2\beta, \\ 
& Y = \mu \, v \, (m_A^2-2\mu^2) + \frac{v^3}{4} \left(g_1^2 M_1 + g_2^2
  M_2\right) \, \sin 4\beta.
\end{aligned}
\end{eqnarray}

\noindent
In writing~\eqref{fields}, we assume following Ref.~\cite{Astapov:2014mea}
that sgoldstino field $\phi$ does not acquire non-zero vacuum
expectation value\footnote{It was shown in~\cite{Astapov:2014mea}, that the
  third derivatives of sgoldstino K{\"a}hler potential can be adjusted
  in such a way that $\langle \phi \rangle = 0$. This condition can be
  relaxed to a certain extent: non-zero vev of $\phi$ affects
  sgoldstino-Higgs mixing in the order $\frac{1}{F^2}$ and thus
  suppressed as compared to the leading contribution if $\langle \phi
  \rangle \lsim \sqrt{F}$. }. In this study, we address sufficiently
small sgoldstino masses, hence the heavier Higgs boson $H$ decouples
and the remaining light states can be approximated by the following
linear combination
\begin{equation}
\label{mixing1}
\left( \begin{matrix}
\tilde{h} \\ \tilde{s}
\end{matrix} \right) = 
\left( \begin{matrix}
\cos \, \theta & -\sin \, \theta \\
\sin \, \theta & \cos \, \theta
\end{matrix} \right) 
\left( \begin{matrix}
h \\ s
\end{matrix} \right).
\end{equation} 
The mixing angle can be obtained from the following equation
\begin{equation}
\label{mixing2}
\tan 2\theta = \frac{2X}{F(m_s^2-m_h^2)}.
\end{equation}

\noindent
In the decoupling regime, the $m_h^2$ parameter is given by
\begin{equation}
\label{mh}
m_h^2 = m_Z^2\cos^2{2\beta} + \frac{v^2}{F^2}\left(B\sin{2\beta} -
2\mu^2\right)^2 + {\rm loop}\;{\rm corrections}.
\end{equation}
Let us note, that the second term in~\eqref{mh} coming from
Eq.~\eqref{correction} at some values of parameters gives considerable
contribution~\cite{Petersson:2012nv} and allows to reduce the level of
fine-tuning as compared to the standard MSSM setup for
$\sqrt{F}\sim$~few TeV, see~\cite{Antoniadis:2014eta} for details.  

\noindent
The mixing between  the Higgs boson and sgoldstino results in
modification of their coupling constants with the vector bosons and SM 
fermions. It is important for our study, that sgoldstino interactions
with leptons are given by the soft trilinear couplings $A^L_{ab}$. In
a generic model, their flavor structure is different from that of
lepton Yukawa coupling constants. In this way, small admixture of
sgoldstino to the lightest Higgs boson generates flavor violating
couplings of the latter. To describe changes of the couplings, let us
consider the relevant part of the lagrangian after the EWSB 
\begin{eqnarray}
\label{no8}
\begin{aligned}
& \mathcal{L} \supset Y^{L}_{ab} \, \widebar{e}_b \, l_a
  \left(1+\frac{h}{\sqrt{2}v}\right) + \frac{A^L_{ab}}{\sqrt{2}F}
  \widebar{e}_b \, l_a \, v_d \, s + \text{h.c.} \supset \\  
& \supset (v_d \, \widebar{l_R} \, Y^L \, l_L + \text{h.c.}) + \left(v_d \, \widebar{l_R} \left(\frac{Y^L}{\sqrt{2}v} \cos \theta -
  \frac{(A^L)^T}{\sqrt{2}F} \sin \theta \right) l_L \tilde{h} +
  \text{h.c.} \right).
\end{aligned}
\end{eqnarray}
Assuming the leptons to be in the mass basis $l = \left(e, \mu,
\tau\right)^{\rm T}$ with $v_d Y_{ab}^L = -m_a \delta_{ab}$,
we obtain 
\begin{equation}
\label{no9}
\mathcal{L} \supset -m_a \, \widebar{l^a} \, l^a-\tilde{h}\left(\widebar{l^a_L} \,
l_R^b \, Y^{\tilde{h}}_{ab} + \text{h.c.} \right),  
\end{equation}
where  the modified Yukawa couplings look as
\begin{equation}
\label{no10}
Y^{\tilde{h}}_{ab} = \frac{m_a \, \delta_{ab} \, \cos \theta}{\sqrt{2}v} + \frac{v_d A^L_{ab} \, \sin \theta}{\sqrt{2}F}.
\end{equation}
We see that  $Y^{\tilde{h}}_{ab} \neq 0$ if $a \neq b$ and hence the LVF decays of the Higgs boson arise already at tree level. The decay width for $\tilde{h} \rightarrow l_a \, l_b$ with $a \neq b$ is given by~\cite{Khachatryan:2015kon,Harnik:2012pb}  
\begin{equation}
\label{no12}
\Gamma(\tilde{h} \rightarrow l_a \, l_b) = \Gamma(\tilde{h}
\rightarrow \widebar{l}_b \, l_a) + \Gamma(\tilde{h} \rightarrow \widebar{l}_a
\, l_b) = \frac{m_{\tilde{h}}}{8\pi} \left(\vert Y^{\tilde{h}}_{ab}\vert^2+\vert
Y^{\tilde{h}}_{ba} \vert^2 \right).
\end{equation}

\section{Analysis of the scenario}
In this section, we describe the strategy which is used here to analyze
phenomenological consequences of the scenario with lepton
flavor-violating couplings of the Higgs resonance which appear from
its interactions with the sector responsible for supersymmetry
breaking. Although this scenario is quite general and allows for
flavor violation in both quark and lepton sectors, in the following
discussion we focus mainly on $\mu$--$\tau$ part in view 
of the CMS excess. We perform a scan over relevant part of the
parameter space presented in Table~\ref{tbl2}.
\begin{table}[h!]
\begin{center}
\begin{tabular}{|M{3cm}|M{8cm}|N}
\hline
$\tan \, \beta$ & 1.5 \ldots 50.5 & \\[4pt] \hline
$|\mu|$ & $100$ \ldots $2000$ $\text{GeV}$ & \\[4pt] \hline
$M_1$ & 100 $ \ldots 2000 \, \text{GeV}$ & \\[4pt] \hline
$M_2$ & 200 $ \ldots 2000 \, \text{GeV}$ & \\[4pt] \hline
$M_3$ & 1.5  $\ldots 4.0 \, \text{TeV}$ & \\[4pt] \hline
$A_{\tau \tau},  A_{\mu \tau}, A_{\tau \mu}$ & 0.1 $\sqrt{F}$ $\ldots$ $\sqrt{F}$ & \\[4pt] \hline
$A_{\mu \mu}$ & 0.1 $\sqrt{F}$ $\ldots$ $0.5 \sqrt{F}$ & \\[4pt] \hline
$ m_{\text{sl}}$  & 4000~$\text{GeV}$ $\ldots \sqrt{F}\, $ & \\[4pt] \hline
\end{tabular}
\end{center}
\caption{Parameter space used in the analysis}
\label{tbl2}
\end{table}
We remind that the consistency of the effective field theory approach
to the model~\eqref{Kahler}--\eqref{SuperpF} requires that the
parameters which become soft terms after the spontaneous supersymmetry breaking should be smaller than $\sqrt{F}$. In what follows, we fix value of supersymmetry breaking scale to 8 TeV. We will comment on this choice later on. Note that we allow for rather large values of off-diagonal trilinear soft parameters $A_{\mu \mu}, A_{\tau \tau}, A_{\mu\tau}$
and $A_{\tau\mu}$ and following purely phenomenological approach
assume no other sources of lepton flavor violation in the model. 
All soft masses of sleptons are chosen to be
equal and we scan over their common value $m_{\text{sl}}$. 
While scanning over the soft parameters of the lepton sector, we take into account experimental constraint on slepton masses. Namely, we will require that the mass of the lightest slepton should be
larger than  $325 \, \text{GeV}$~\cite{Aad:2014vma}. In our analysis, we calculate spectrum of the lepton mass matrix and check whether this constraint on the smallest eigenvalue is fulfilled.   
The squark sector of the model is not considered here, and thus we
independently scan over the mass parameter $m_{h}$ of the lightest
Higgs boson entering the scalar mass matrix over the following
interval $115$--$130$~GeV. We find that for the most interesting cases
the mass parameter of the scalar sgoldstino should not be very heavy
or very small. In the case of heavy sgoldstino, the mixing
angle~\eqref{mixing2} is  small and, as a consequence, the width of $\tilde{h}\to\mu\tau$ decay is suppressed. On the other hand, very light sgoldstinos with large Higgs boson admixture are phenomenologically unacceptable due to
results from the LEP~\cite{Schael:2006cr} and
Tevatron~\cite{Abazov:2013gmz} experiments. In what follows, we limit  
ourselves to the regimes in which the scalar sgoldstino mass parameter
is somewhat smaller (90--114~GeV) or larger (150--200~GeV) than the Higgs boson mass. Some parameters, which are not of primary importance for the analysis, were fixed to reasonable
benchmark values.  In particular, we set the soft trilinear constant
of b-quark $A_{bb} = 0.5 \,\sqrt{F}$ and the mass of pseudoscalar
sgoldstino $m_{p}=200\, \text{GeV}$. For each chosen point in the
parameter space, we find physical masses of the Higgs-like
$m_{\tilde{h}}$ and sgoldstino-like states $m_{\tilde{s}}$, selecting
models with the Higgs resonance lying in the mass range $m_{\tilde{h}}
= 125.09^{+0.24}_{-0.24}~\rm{GeV}$, calculate relevant observables
which will be discussed below and find 
phenomenologically acceptable models.     

\noindent
Mixing of scalar sgoldstino with the lightest Higgs boson results in
modifications of the Higgs signal strengths 
\begin{equation}
\label{signal}
\mu_f = \frac{\sigma(pp \rightarrow \tilde{h}) \times
 {\rm Br}(\tilde{h} \rightarrow f)}{\sigma(pp \rightarrow h^{SM})
  \times {\rm Br}(h^{SM} \rightarrow f)}, 
\end{equation}
where index $f$ stands for the following final states, $W^+ \, W^-$,
$ZZ$, $\gamma \, \gamma$,  $\widebar{b} \, b$, $\tau^+ \, \tau^-$ and $\mu^+\mu^-$. We calculate them using modified Higgs 
boson couplings presented in Appendix~A. Sizable QCD corrections have been taken into  account using general expressions from Ref.~\cite{Spira:1997dg}. For the diboson final states, the Higgs boson is mainly produced at the LHC via gluon-gluon fusion 
(\text{ggh})  channel  and neglecting other production mechanisms is a fairly good approximation. In this case, one obtains 
\begin{equation}
\label{ggH}
\frac{\sigma(pp \rightarrow \tilde{h})}{\sigma (pp \rightarrow
  h)_{\text{SM}}} \simeq \frac{\Gamma(\tilde{h} \rightarrow
  gg)}{\Gamma(h \rightarrow gg)_{SM}} = \frac{\left \vert \sum_{Q}
  A_{1/2}(\tau_Q) \, \cos \theta + \frac{6 M_3 \pi v}{\alpha_s \, F} \,
  \sin \theta \right \vert^2}{\left \vert \sum_{Q} A_{1/2}(\tau_Q)
  \right \vert^2}\;,
\end{equation}
where the sum goes over all quarks; see Appendix~A.
It should be noticed, that both terms in the numerator, $\sum_{Q}
A_{1/2}(\tau_Q)$ and $\frac{6 M_3 \pi v}{\alpha_s \, F}$, can be of
the same size. So, in the case when $M_3$ and $\sin \theta$ have
different  signs (for example in case of the negative value of the
parameter $\mu$ and positive $M_3$) the ratio~\eqref{ggH} can be close
to unity even in the case of large mixing angle. This possibility can
provide with sizable off-diagonal Yukawas $Y_{\mu \tau}^{\tilde{h}}$
($Y_{\tau \mu}^{\tilde{h}}$) and fairly large branching of process
$\tilde{h} \rightarrow \mu \tau$ (see discussion section).  In
Table~\ref{tbl1} we present experimental  bounds on $\mu_f$ from the
ATLAS and CMS experiments
\begin{table}[h!]
\begin{center}
  \begin{tabular}{|M{2cm}|M{4.7cm}|M{3.3cm}|M{3.4cm}|N}
\hline
\textbf{Decay channel} & \textbf{Production channel used in the
  analysis} & \textbf{$\mu_f$, CMS} & \textbf{$\mu_f$, ATLAS}
\\ \hline 
$\tilde{h} \rightarrow b \widebar{b}$ & production in association
\newline with a vector boson (\text{Vh}) & $0.89 \pm 0.43$
\cite{Chatrchyan:2013zna,Khachatryan:2015bnx,Chatrchyan:2014vua} &
$0.74^{+0.17}_{-0.16}$ \cite{Aad:2014xzb} \\ \hline 
$\tilde{h} \rightarrow \tau \widebar{\tau}$ & gluon-gluon fusion
(\text{ggh}), vector-boson fusion (\text{VBF}), associated
production (\text{Vh}) & $0.94 \pm 0.41$ -- \text{VBF} 
\newline $1.07 \pm 0.46$ -- \text{ggh} 
\newline \cite{Chatrchyan:2014nva,Chatrchyan:2014vua} & $1.4 \pm 0.4$ --
\text{VBF} \cite{Aad:2015vsa} 
\\ \hline 
$\tilde{h} \rightarrow WW$ & gluon-gluon fusion (\text{ggh}) &
$0.74^{+0.22}_{-0.20}$ \cite{Chatrchyan:2013iaa} & $1.02^{+0.29}_{-0.26}$
\cite{ATLAS:2014aga} & \\[10pt] \hline 
$\tilde{h} \rightarrow ZZ$ & gluon-gluon fusion (\text{ggh}),
vector-boson fusion (\text{VBF}), associated production
(\text{Vh}), quarks-fusion (\text{tth}, \text{bbh}) &
$0.83^{+0.31}_{-0.25}$ \cite{Chatrchyan:2013mxa} & $1.44^{+0.40}_{-0.33}$
\cite{Aad:2014eva} \\ \hline 
$\tilde{h} \rightarrow \gamma \gamma$ & gluon-gluon fusion
(\text{ggh}) & $1.12^{+0.37}_{-0.32}$ \cite{Khachatryan:2014ira} & $1.32 \pm
0.38$ \cite{Aad:2014eha} &\\[10pt] \hline 
\end{tabular}
\caption {Constraints on the signal strengths $\mu_f$ from the LHC
  experiments} 
\label{tbl1}
\end{center}
\end{table}
for different production and decay channels of Higgs boson which are
taken into account in the present analysis. We accept given point in
parameter space (see below) if it predicts $\mu_f$ which lies inside
the ATLAS and CMS bounds. Mixing of the Higgs boson with
sgoldstino leads to significant modification of its decay into a pair of muons in comparison with the SM. This decay has not been seen yet at the LHC. The best upper limits for its signal strengths are $\mu_{\mu \mu} < 13.2$ for ggH production
channel and $\mu_{\mu \mu} < 11.2$ for VBF channel \cite{Khachatryan:2014aep}. The ratio of the corresponding decay
widths in our model and in the SM for Higgs bosons of the same mass
reads  
\begin{equation}
\label{widthmumu}
\frac{\Gamma(\tilde{h}\to \mu \mu)}{\Gamma (h \to \mu
  \mu)}_{\text{SM}} = \left(\cos \theta + \frac{A_{\mu\mu} v^2}{F
  m_{\mu}} \cos \beta \sin \theta \right)^2.
\end{equation}
For $A_{\mu\mu} \sim \sqrt{F}$ this ratio is large and exceeds
the experimental bounds. For this reason, we choose the upper bound	
for $A_{\mu \mu}$ in our scanning to be equal to $0.5 \sqrt{F}$ 
(see Table \ref{tbl2}). 

\noindent
Further, we check whether the scalar sgoldstino-like resonance is allowed 
by existing experimental results. For the case of the LHC searches for
diboson resonances, we use the observables $\sigma(pp \rightarrow
\tilde{s}) \times \text{Br} (\tilde{s} \rightarrow f)$ 
where $f$ stands for pair of photons~\cite{Khachatryan:2015qba}, 
$W$~\cite{Aad:2012uub,Khachatryan:2015cwa} or $Z$
bosons~\cite{Aad:2015kna}. These final states are the most constraining
for sgoldstino with discussed parameters. Due to large tree level
couplings to  the massless vector bosons dominating production
mechanism 
for sgoldstino will be gluon-gluon fusion~\cite{Perazzi:2000ty}. The
leading order production cross section can be written in the form 
\begin{equation}
\label{no27}
\sigma_{\tilde{s}} = \frac{\pi^2}{8} \frac{\Gamma(\tilde{s}\rightarrow 
  gg)}{s m_{\tilde{s}}} \int_{m_{\tilde{s}}^2/s}^{1} \frac{dx}{x} \,
f_{p/g}(x,m_{\tilde{s}^2}) \,
f_{p,\widebar{p}/g}\left(\frac{m_{\tilde{s}}^2}{xs},m_{\tilde{s}}^2\right), 
\end{equation}
where $\Gamma(\tilde{s}\rightarrow gg)$ is the partial width of
sgoldstino-like state decaying into two gluons, $s$ is the center of
mass energy squared and $f_{p/g}(x,Q^2)$ are the parton distribution
functions defined at scale $Q^2$. We numerically calculate the
quantity  $\sigma_{\tilde{s}} \times \text{BR}(\tilde{s} \rightarrow \gamma \gamma)$ using \texttt{CTEQ6L} ~\cite{Gao:2013xoa} parametrization of the parton distribution functions and compare it with the experimental bounds. Very light sgoldstino $\tilde{s}$, with the mass in the range 90--114~GeV and with sufficiently large Higgs boson admixture, decays 
dominantly into $b\bar{b}$ final state. We use corresponding bounds
from LEP~\cite{Schael:2006cr} and TeVatron~\cite{Abazov:2013gmz} 
searches in this case. 

\noindent
Now, let us turn to the observables specific for lepton flavor
violation in question. Interactions of the Higgs boson $\tilde{h}$ and scalar sgoldstino $\tilde{s}$ in $\mu$--$\tau$ sector are described by the following lagrangian    
\begin{eqnarray}
\label{LFVLagr}
\begin{aligned}
& \mathcal{L} \supset - Y_{\mu \tau}^{\tilde{h}} \, \tilde{h} \,
  \widebar{\mu}_L\tau_R - Y_{\tau \mu}^{\tilde{h}} \, \tilde{h} \,
  \widebar{\tau}_L \mu_R - Y_{\mu \mu}^{\tilde{h}} \, \tilde{h} \,
  \widebar{\mu}_L \mu_R -Y_{\tau \tau}^{\tilde{h}} \, \tilde{h} \, 
  \widebar{\tau}_L \tau_R - \\ 
& - Y_{\mu \tau}^{\tilde{s}} \, \tilde{s} \, \widebar{\mu}_L \tau_R -
  Y_{\tau \mu}^{\tilde{s}} \, \tilde{s} \, \widebar{\tau}_L \mu_R -
  Y_{\mu \mu}^{\tilde{s}} \, \tilde{s} \, \widebar{\mu}_L \mu_R -
  Y_{\tau \tau}^{\tilde{s}} \, \tilde{s} \, \widebar{\tau}_L \tau_R  + 
  \text{h.c.} 
\end{aligned}
\end{eqnarray}
These interactions contribute several lepton flavor-violating
processes. For our analysis, the most important of
them
are
$\tau \rightarrow \mu \gamma$ and  $\tau \rightarrow 3\mu$
decays. 
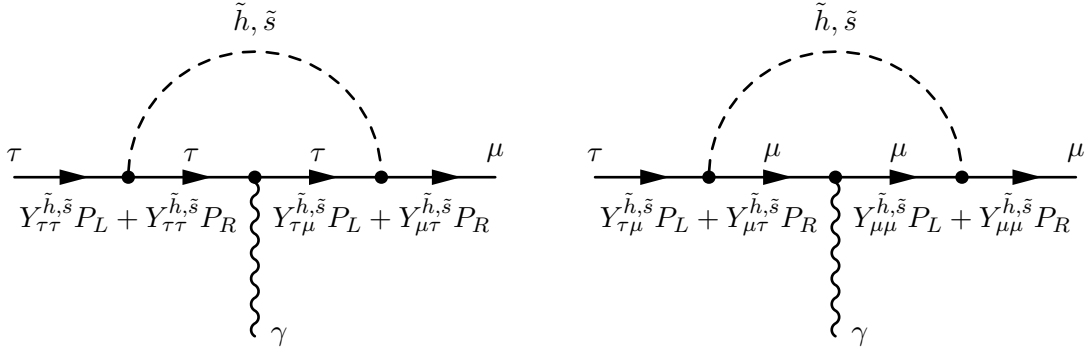
\begin{figure}[!htb]
\centering
\begin{fmffile}{1-loop-h}
\begin{fmfgraph*}(180,120)
\fmfleft{i}
\fmfright{o2}
\fmfbottom{o1}
\fmf{fermion,tension=2}{i,v1}
\fmf{fermion,tension=2}{v3,o2}
\fmf{dashes,left,label=$\tilde{h},,\tilde{s}$,tension=0.4}{v1,v3}
\fmf{fermion,label=$\tau$,label.side=left}{v1,v2}
\fmf{fermion,label=$\tau$,label.side=left}{v2,v3}
\fmfdot{v1,v2,v3}
\fmf{photon,tension=0}{v2,o1}

\fmfv{label=$\tau$,label.angle=90}{i}
\fmfv{label=$\mu$,label.angle=90}{o2}
\fmfv{label=$\gamma$,label.angle=0}{o1}
\fmfv{label=$Y_{\tau \tau}^{\tilde{h},,\tilde{s}} P_L+Y_{\tau
    \tau}^{\tilde{h},,\tilde{s}} P_R$,label.angle=-90, 
decor.shape=circle,
decor.filled=full,decor.size=2thick}{v1}
\fmfv{label=$Y_{\tau \mu}^{\tilde{h},,\tilde{s}} P_L+Y_{\mu
    \tau}^{\tilde{h},,\tilde{s}} P_R$,label.angle=-90, 
decor.shape=circle,
decor.filled=full,decor.size=2thick}{v3}
\end{fmfgraph*}
\hspace{30px}
\begin{fmfgraph*}(180,120)
\fmfleft{i}
\fmfright{o2}
\fmfbottom{o1}
\fmf{fermion,tension=2}{i,v1}
\fmf{fermion,tension=2}{v3,o2}
\fmf{dashes,left,label=$\tilde{h},,\tilde{s}$,tension=0.4}{v1,v3}
\fmf{fermion,label=$\mu$,label.side=left}{v1,v2}
\fmf{fermion,label=$\mu$,label.side=left}{v2,v3}
\fmfdot{v1,v2,v3}
\fmf{photon,tension=0}{v2,o1}

\fmfv{label=$\tau$,label.angle=90}{i}
\fmfv{label=$\mu$,label.angle=90}{o2}
\fmfv{label=$\gamma$,label.angle=0}{o1}
\fmfv{label=$Y_{\tau \mu}^{\tilde{h},,\tilde{s}} P_L+Y_{\mu
    \tau}^{\tilde{h},,\tilde{s}} P_R$,label.angle=-90, 
decor.shape=circle,
decor.filled=full,decor.size=2thick}{v1}
\fmfv{label=$Y_{\mu \mu}^{\tilde{h},,\tilde{s}} P_L+Y_{\mu \mu}^{\tilde{h},,\tilde{s}} P_R$,label.angle=-90, 
decor.shape=circle,
decor.filled=full,decor.size=2thick}{v3}
\end{fmfgraph*}
\end{fmffile}
\caption{1-loop diagrams with $\tilde{h}$ and $\tilde{s}$}
\label{fig:1lh}
\end{figure}
\noindent
The effective lagrangian describing $\tau \rightarrow \mu \gamma$
decay is 
\begin{equation}
\label{Lefftmg}
\mathcal{L}_{\text{eff}} = c_L \, \frac{e}{8\pi^2} \, m_{\tau}
\left (\widebar{\mu}\, \sigma^{\alpha \beta} \, P_L \tau \right) \, F_{\alpha \beta} +
c_R \,  \frac{e}{8\pi^2} \, m_{\tau} \left( \widebar{\mu} \, \sigma^{\alpha
  \beta} \, P_R \tau \right) \, F_{\alpha \beta} +\text{h.c.} 
\end{equation}
and the corresponding decay width is given by
\begin{equation}
\label{tmgWidth}
\Gamma(\tau \rightarrow \mu \gamma) = \frac{\alpha \, m_{\tau}^5}{64
  \pi^4} \left(\vert c_L \vert^2+\vert c_R \vert^2 \right).
\end{equation}
Here $c_{L,R}$ are the Wilson coefficients which acquire different
contributions from the standard diagrams involving sleptons as well
as model-specific contributions from loop diagrams containing the
Higgs boson and sgoldstino with lepton flavor-violating
couplings~\eqref{Lefftmg}. Now we remind that the effective theory
with spontaneous supersymmetry breaking which we consider in this
paper is not renormalizable and the one-loop contribution of goldstino
sector to  the coefficients $c_{L,R}$ is in general divergent. In
fact, one can write higher dimensional supersymmetric operator which
would generate the terms as in eq.~\eqref{Lefftmg} already at tree
level after supersymmetry breaking (see,
e.g. Ref.~\cite{Brignole:1999gf}). In this sense, our model has a
limited predictive power with respect to such observables as ${\rm
  Br}(\tau\to\mu\gamma)$ or ${\rm Br}(\tau\to 3\mu)$ which depend on
underlying microscopic theory. To have a glimpse on possible size of
the effect, we assume that there is no tree level contribution to the
lagrangian~\eqref{Lefftmg} but it appears at one-loop level. We will estimate the dominant  divergent one-loop contributions assuming a realistic cutoff $\Lambda$ for the effective theory.
Possible values of the cutoff in the low scale supersymmetry breaking models have been discussed some time ago in
Refs.~\cite{Brignole:2000wd,Brignole:1998uu,Brignole:1996fn}. It has been found that the cutoff for this model can lie somewhere between the level of soft masses of matter scalars $\tilde{m}$ (the largest of which can not exceed $\sqrt{F}$) and the value $\Lambda^2 = 16\pi F^2/\tilde{m}^2$. The latter represents the energy at which perturbative unitarity is violated in the model in $2\to 2$ scattering of matter fermions. In our numerical estimates for $\rm{Br}(\tau \to \mu \gamma)$ we use the upper boundary of the allowed region of the cutoffs with $\tilde{m}$ replaced by the level of slepton masses, as the sleptons are most relevant for our analysis. Here we refer interested reader to Refs.~\cite{Brignole:2000wd,Brignole:1999gf} for extensive discussions of loop contributions of goldstino sector to different FCNC processes and muon anomalous magnetic moment. Having made this  disclaimer, we collect different parts of the Wilson coefficients $c_{L,R}$ as follows    
\begin{equation}
\label{WilsonC}
c_{L,R} =
c_{L,R}^{\text{1-loop},\tilde{h}}+c_{L,R}^{\text{1-loop},\tilde{s}} +
c_{L,R}^{\text{2-loop}}+c_{L,R}^{sp} + c_{L,R}^{\text{SUSY}}, 
\end{equation}
where
$c_{L,R}^{\text{1-loop},\tilde{h}},c_{L,R}^{\text{1-loop},\tilde{s}}$
are convergent one-loop contribution with the Higgs boson and 
sgoldstino in Fig.~\ref{fig:1lh}, $c_{L,R}^{\text{2-loop}}$ are 2-loop Barr-Zee type diagrams presented in Fig.~\ref{fig:2lh}
\begin{figure}[!htb]
\centering
\begin{fmffile}{2-loop-01}
\begin{fmfgraph*}(180,120)
\fmfleft{i}
\fmfright{o2}
\fmftop{i,d1,o2}
\fmfbottom{o1}
\fmf{fermion}{i,v1}
\fmf{fermion,label=$\mu$,label.side=right}{v1,v2}
\fmf{fermion}{v2,o2}
\fmffreeze
\fmf{photon,tension=0.6,label= $Z,, \gamma$}{v2,v3}
\fmf{dashes,tension=0.6,label=$\tilde{h},, \tilde{s}$}{v1,v4}
\fmf{fermion,tension=0,label=$t$,label.side=right}{v4,v3}
\fmf{fermion,tension=0.4,label=$t$,label.side=left}{v3,v5}
\fmf{fermion,tension=0.4,label=$t$,label.side=left}{v5,v4}
\fmf{photon}{v5,o1}
\fmfv{label=$\tau$,label.angle=90}{i}
\fmfv{label=$\mu$,label.angle=90}{o2}
\fmfv{label=$\gamma$,label.angle=0}{o1}
\fmfv{decor.shape=circle,decor.filled=full,decor.size=2thick}{v1}
\fmfv{decor.shape=circle,decor.filled=full,decor.size=2thick}{v2}
\fmfv{decor.shape=circle,decor.filled=full,decor.size=2thick}{v3}
\fmfv{decor.shape=circle,decor.filled=full,decor.size=2thick}{v4}
\fmfv{decor.shape=circle,decor.filled=full,decor.size=2thick}{v5}
\end{fmfgraph*}
\hspace{20px}
\begin{fmfgraph*}(180,120)
\fmfleft{i}
\fmfright{o2}
\fmftop{i,d1,o2}
\fmfbottom{o1}
\fmf{fermion}{i,v1}
\fmf{fermion,label=$\mu$,label.side=right}{v1,v2}
\fmf{fermion}{v2,o2}
\fmffreeze
\fmf{photon,tension=0.6,label= $Z,, \gamma$}{v2,v3}
\fmf{dashes,tension=0.6,label=$\tilde{h},,\tilde{s}$}{v1,v4}
\fmf{photon,tension=0,label=$W$,label.side=right}{v4,v3}
\fmf{photon,tension=0.4,label=$W$,label.side=left}{v3,v5}
\fmf{photon,tension=0.4,label=$W$,label.side=left}{v5,v4}
\fmf{photon}{v5,o1}
\fmfv{label=$\tau$,label.angle=90}{i}
\fmfv{label=$\mu$,label.angle=90}{o2}
\fmfv{label=$\gamma$,label.angle=0}{o1}
\fmfv{decor.shape=circle,decor.filled=full,decor.size=2thick}{v1}
\fmfv{decor.shape=circle,decor.filled=full,decor.size=2thick}{v2}
\fmfv{decor.shape=circle,decor.filled=full,decor.size=2thick}{v3}
\fmfv{decor.shape=circle,decor.filled=full,decor.size=2thick}{v4}
\fmfv{decor.shape=circle,decor.filled=full,decor.size=2thick}{v5}
\end{fmfgraph*}
\end{fmffile}
\end{figure}
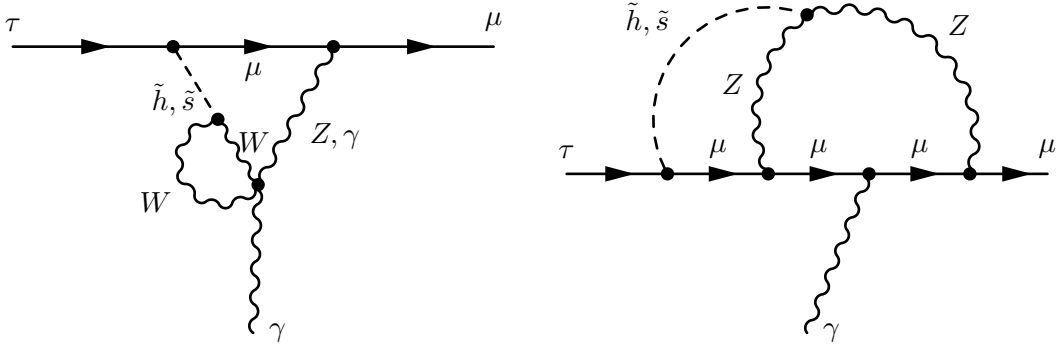
\begin{figure}[!htb]
\centering
\vspace{30px}
\begin{fmffile}{2-loop-02}
\begin{fmfgraph*}(180,120)
\fmfleft{i}
\fmfright{o2}
\fmftop{i,d1,o2}
\fmfbottom{o1}
\fmf{fermion}{i,v1}
\fmf{fermion,label=$\mu$,label.side=right}{v1,v2}
\fmf{fermion}{v2,o2}
\fmffreeze
\fmf{photon,tension=0.6,label= $Z,, \gamma$,label.side=left}{v2,v3}
\fmf{dashes,tension=0.9,label=$\tilde{h},,\tilde{s}$,label.side=right,label.dist=0.1}{v1,v4}
\fmf{photon,tension=0,label=$W$,label.side=left,label.dist=0.01}{v4,v3}
\fmf{photon,right=1.6,label=$W$}{v4,v3}
\fmf{photon}{v3,o1}
\fmfv{label=$\tau$,label.angle=90}{i}
\fmfv{label=$\mu$,label.angle=90}{o2}
\fmfv{label=$\gamma$,label.angle=0}{o1}
\fmfv{decor.shape=circle,decor.filled=full,decor.size=2thick}{v1}
\fmfv{decor.shape=circle,decor.filled=full,decor.size=2thick}{v2}
\fmfv{decor.shape=circle,decor.filled=full,decor.size=2thick}{v3}
\fmfv{decor.shape=circle,decor.filled=full,decor.size=2thick}{v4}
\end{fmfgraph*}
\hspace{20px}
\begin{fmfgraph*}(180,120)
\fmfleft{i}
\fmfright{o2}
\fmftop{d1}
\fmfbottom{o1}
\fmf{fermion}{i,v1}
\fmf{fermion,label=$\mu$,label.side=left}{v1,v2}
\fmf{fermion,label=$\mu$,label.side=left}{v2,v3}
\fmf{fermion,label=$\mu$,label.side=left}{v3,v4}
\fmf{fermion,tension=1.3}{v4,o2}
\fmffreeze
\fmf{dashes,left=0.7,label=$\tilde{h},,\tilde{s}$}{v1,d1}
\fmf{photon,left=0.35,label=$Z$,tension=0.4}{v2,d1}
\fmf{photon,left=0.6,label=$Z$,tension=0.4}{d1,v4}
\fmf{photon,tension=0}{v3,o1}
\fmfv{label=$\tau$,label.angle=90}{i}
\fmfv{label=$\mu$,label.angle=90}{o2}
\fmfv{label=$\gamma$,label.angle=0}{o1}
\fmfv{decor.shape=circle,decor.filled=full,decor.size=2thick}{v1}
\fmfv{decor.shape=circle,decor.filled=full,decor.size=2thick}{v2}
\fmfv{decor.shape=circle,decor.filled=full,decor.size=2thick}{v3}
\fmfv{decor.shape=circle,decor.filled=full,decor.size=2thick}{v4}
\fmfv{decor.shape=circle,decor.filled=full,decor.size=2thick}{d1}
\end{fmfgraph*}
\end{fmffile}
\caption{Barr-Zee type 2-loop diagrams}
\label{fig:2lh}
\end{figure}
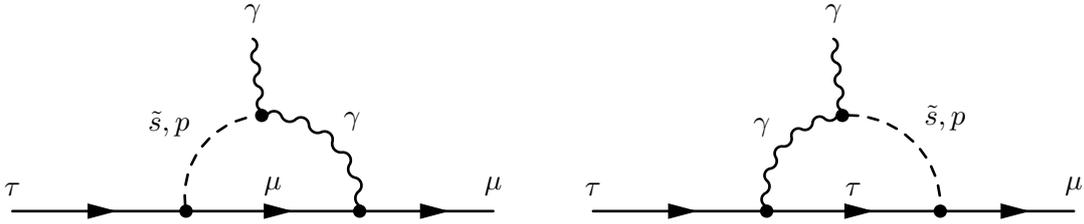
\begin{figure}[h!]
\centering
\begin{fmffile}{1-loop-sp}
\vspace{30px}
\begin{fmfgraph*}(180,130)
\fmfleft{i}
\fmfright{o2}
\fmftop{d0}
\fmftop{d1}
\fmf{fermion}{i,v2}
\fmf{fermion,label=$\mu$,label.side=left}{v2,v4}
\fmf{fermion,tension=1.3}{v4,o2}
\fmffreeze
\fmf{dashes,left=0.45,label=$\tilde{s},,p$,tension=0.4}{v2,v1}
\fmf{photon,left=0.45,label=$\gamma$,tension=0.4}{v1,v4}
\fmf{photon,tension = 0.0}{v1,d1}
\fmfv{label=$\tau$,label.angle=90}{i}
\fmfv{label=$\mu$,label.angle=90}{o2}
\fmfv{label=$\gamma$}{d1}
\fmfv{decor.shape=circle,decor.filled=full,decor.size=2thick}{v1}
\fmfv{decor.shape=circle,decor.filled=full,decor.size=2thick}{v2}
\fmfv{decor.shape=circle,decor.filled=full,decor.size=2thick}{v4}
\end{fmfgraph*}
\hspace{30px}
\begin{fmfgraph*}(180,130)
\fmfleft{i}
\fmfright{o2}
\fmftop{d1}
\fmf{fermion}{i,v2}
\fmf{fermion,label=$\tau$,label.side=left}{v2,v4}
\fmf{fermion,tension=1.3}{v4,o2}
\fmffreeze
\fmf{photon,left=0.45,label=$\gamma$,tension=0.4}{v2,v1}
\fmf{dashes,left=0.45,label=$\tilde{s},,p$,tension=0.4}{v1,v4}
\fmf{photon,tension = 0.0}{v1,d1}
\fmfv{label=$\tau$,label.angle=90}{i}
\fmfv{label=$\mu$,label.angle=90}{o2}
\fmfv{label=$\gamma$}{d1}
\fmfv{decor.shape=circle,decor.filled=full,decor.size=2thick}{v1}
\fmfv{decor.shape=circle,decor.filled=full,decor.size=2thick}{v2}
\fmfv{decor.shape=circle,decor.filled=full,decor.size=2thick}{v4}
\end{fmfgraph*}
\end{fmffile}
\caption{1-loop diagrams with internal (pseudo)scalar sgoldstino}
\label{sploops}
\end{figure}
$c_{L.R}^{sp}$ are the one-loop divergent diagrams involving
sgoldstino coupling with photons shown in Fig.~\ref{sploops} and
$c_{L,R}^{SUSY}$ are the 1-loop diagrams with internal superpartners
depicted in Fig.~\ref{fig:lsusy}. 
\begin{figure}[!h]
\centering
\begin{fmffile}{1-loop-susy}
\begin{fmfgraph*}(180,130)
\fmfleft{i}
\fmfright{o2}
\fmftop{d1}
\fmftop{d2}
\fmf{fermion}{i,v2}
\fmf{photon,label=$\tilde{N}$,label.side=right,tension=0.5}{v2,v4}
\fmf{plain,label=$\tilde{N}$,label.side=right,tension=0.5}{v2,v4}
\fmf{fermion,tension=1.3}{v4,o2}
\fmffreeze
\fmf{dashes,left=0.45,label=$\tilde{l}$,label.side=left,tension=0.4}{v2,v1}
\fmf{dashes,left=0.45,label=$\tilde{l}$,label.side=left,tension=0.4}{v1,v4}
\fmf{photon,tension = 0.0}{v1,d1}
\fmfv{label=$\tau$,label.angle=90}{i}
\fmfv{label=$\mu$,label.angle=90}{o2}
\fmfv{label=$\gamma$}{d1}
\fmfv{decor.shape=circle,decor.filled=full,decor.size=2thick}{v1}
\fmfv{decor.shape=circle,decor.filled=full,decor.size=2thick}{v2}
\fmfv{decor.shape=circle,decor.filled=full,decor.size=2thick}{v4}
\end{fmfgraph*}
\end{fmffile}
\caption{1-loop diagram with internal neutralino and sleptons}
\label{fig:lsusy}
\end{figure}
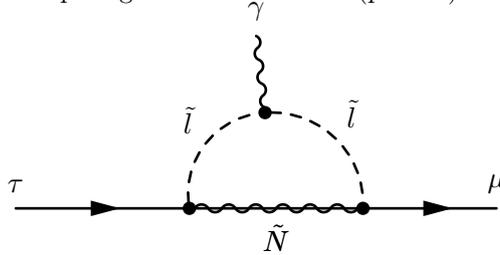
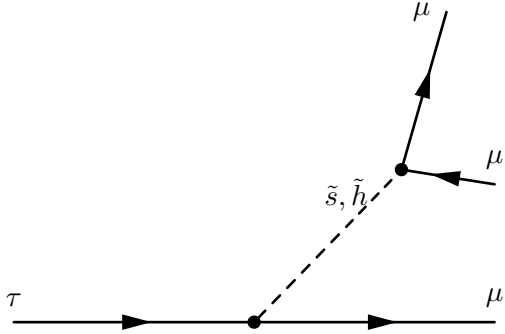
\begin{figure}[h!]
\centering
\begin{fmffile}{tmmm}
\begin{fmfgraph*}(180,130)
\fmfbottom{i1,d1,o1}
\fmfright{o0,o2,o3}
\fmftop{d3}
\fmftop{d4}	
\fmf{fermion}{i1,v1}
\fmf{fermion}{v1,o1}
\fmffreeze
\fmf{dashes,label = $\tilde{s},,\tilde{h}$,tension=1.5,label.angle=90}{v1,v0}
\fmf{dashes,label = $\tilde{s},,\tilde{h}$,tension=2.5,label.angle=90}{v0,v2}
\fmf{fermion}{o2,v2}
\fmf{fermion}{v2,o3}
\fmfv{label=$\tau$,label.angle=90}{i1}
\fmfv{label=$\mu$,label.angle=90}{o1}
\fmfv{label=$\mu$,label.angle=90}{o2}
\fmfv{label=$\mu$,label.angle=180}{o3}
\fmfv{label=$\tilde{s},,\tilde{h}$,label.angle=90}{v0}
\fmfv{decor.shape=circle,decor.filled=full,decor.size=2thick}{v1}
\fmfv{decor.shape=circle,decor.filled=full,decor.size=2thick}{v2}
\end{fmfgraph*}
\end{fmffile}
\caption{Tree-level diagram of $\tau \rightarrow 3\mu$ decay with
  virtual sgoldstino and Higgs exchange} 
\label{fig:tmmm}
\end{figure}
Explicit expressions for
different contributions are presented in Appendix~B. Numerically, we
observe that the dominant contribution for most of the acceptable
models with realistic value of the cutoff $\Lambda$ of microscopic
theory comes from the last term in~\eqref{WilsonC}. We calculate and
sum up different contributions using formulas
(\ref{tmgWidth}),(\ref{C1looph}),(\ref{C2looptg}),(\ref{1loopsusy}),  
find the branching ratio of $\tau\to\mu\gamma$ and compare it with the present $90\%$ C.L. upper limit~\cite{Aubert:2009ag} 
\begin{equation}
\label{BRbound}
{\rm Br}(\tau \to \mu \gamma) < 4.4 \times 10^{-8}.
\end{equation}
Another relevant constraint we use in our analysis is the upper limit on the decay $\tau \to 3\mu$~\cite{Hayasaka:2010np}
\begin{equation}
\label{BRbound2}
{\rm Br}(\tau \to 3 \mu) < 2.1 \times 10^{-8}.
\end{equation}  
The leading order contribution is given by the diagram with exchange
of virtual sgoldstino depicted in Fig.~\ref{fig:tmmm} and reads as
follows
\begin{equation}
\label{t3m}
\Gamma(\tau \to 3\mu) = 
\frac{m_{\tau}^5}{6144 \pi^3} \frac{A_{\mu\mu}^2}{F} \left(\frac{v^2
  \cos^2 \beta}{F}\right)^2
\left(\frac{\cos^2 \theta}{m_{\tilde{s}}^2}+\frac{\sin^2 \theta}{m_{\tilde{h}}^2}\right)^2 \left(\frac{A_{\mu \tau }^2 +
  A_{\tau \mu}^2}{2F} \right). 
\end{equation}
Here we set the mass of muon to zero and contracted the scalar
propagator into point.
Loop corrections to this expression come from diagrams with internal
sfermions and gravitinos~\cite{Brignole:2000wd} and are 
logarithmically divergent. As in the case of $\tau \to \mu \gamma$
decay this reduces predictive power of our model. Estimates with
finite cutoff $\Lambda$ show that this correction is suppressed at
least by the factor $\sim \frac{m_{\tilde{s}}^2}{F} \, \log
\frac{\Lambda}{m_{\tilde{sl}}^2} =  \frac{m_{\tilde{s}}^2}{F} \, \log \frac{16 \pi F^2}{m_{\tilde{sl}}^4}$ as compared to the tree-level contribution. For our
  choice of the parameter space and SUSY breaking scale $\log \frac{16
    \pi F^2}{m_{\tilde{sl}}^4} \lesssim 10$ and the overall
  suppression factor is at least $\lesssim 10^{-2}$. The situation
  changes drastically if one allows for flavour violation  in
  $M_{\tilde{l}LL}^2$ or $M_{\tilde{l}RR}^2$ (see discussion in
  Appendix B). In this case quadratically divergent diagrams come
  into play~\cite{Brignole:2000wd} and more involved analysis is
  needed to obtain precise prediction for $\tau \to 3\mu$. However,   
we are justified to consider tree-level prediction of $\tau \to 3\mu$
as reliable as long as we use the assumptions of our analysis: a) no
flavour violation in $M_{\tilde{l}LL}^2$ or  $M_{\tilde{l}RR}^2$; b)
sgoldstino masses are considerably smaller than SUSY breaking scale; c)
sufficiently large slepton mass scale $m_{\text{sl}}$ (which provides
logarithmic factor of order $10$ in the worst case).

\section{Results and discussion}
\begin{figure}[!htb]
\begin{tabular}{cc}
\includegraphics[width=0.45\columnwidth]{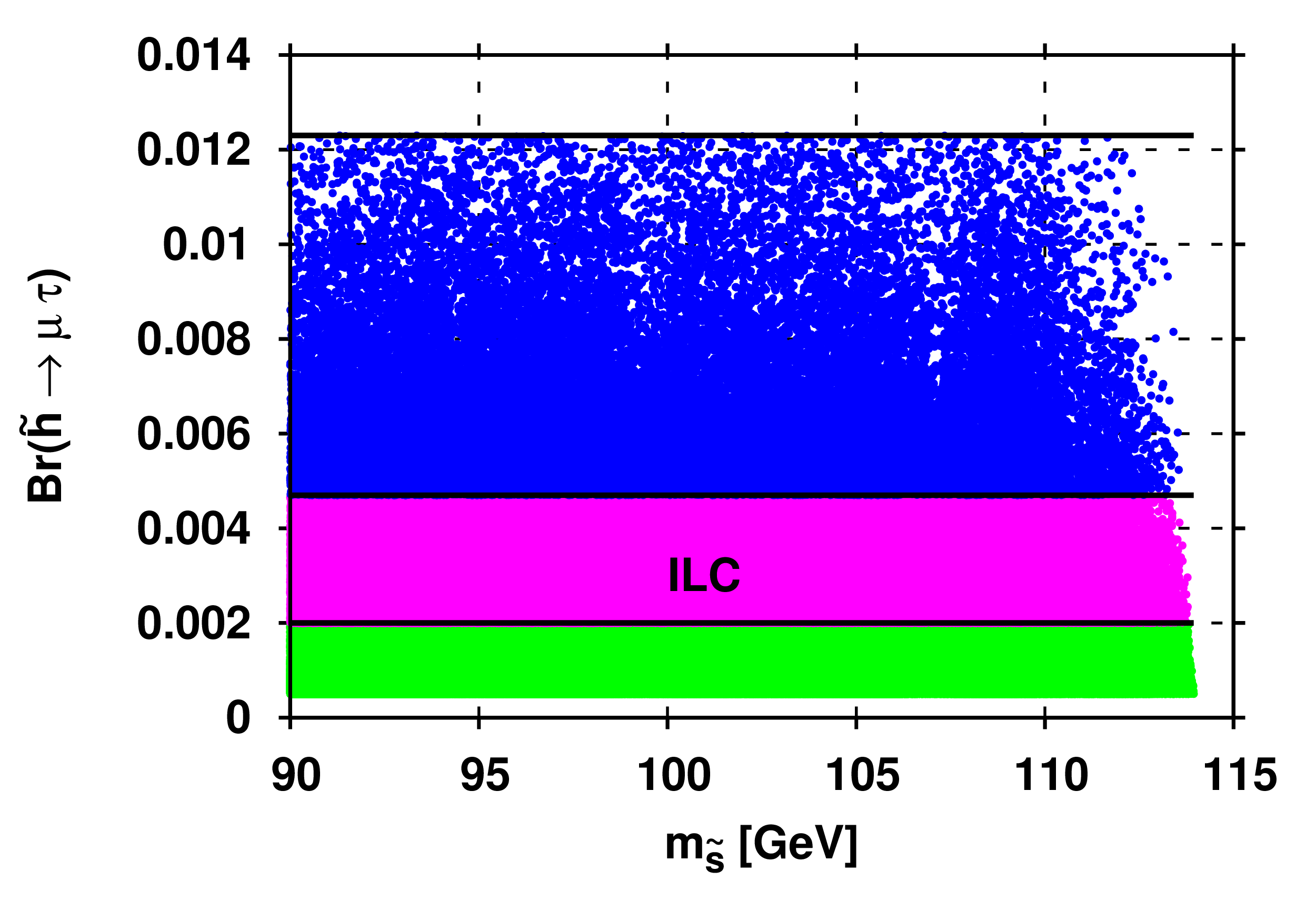}
&
\includegraphics[width=0.45\columnwidth]{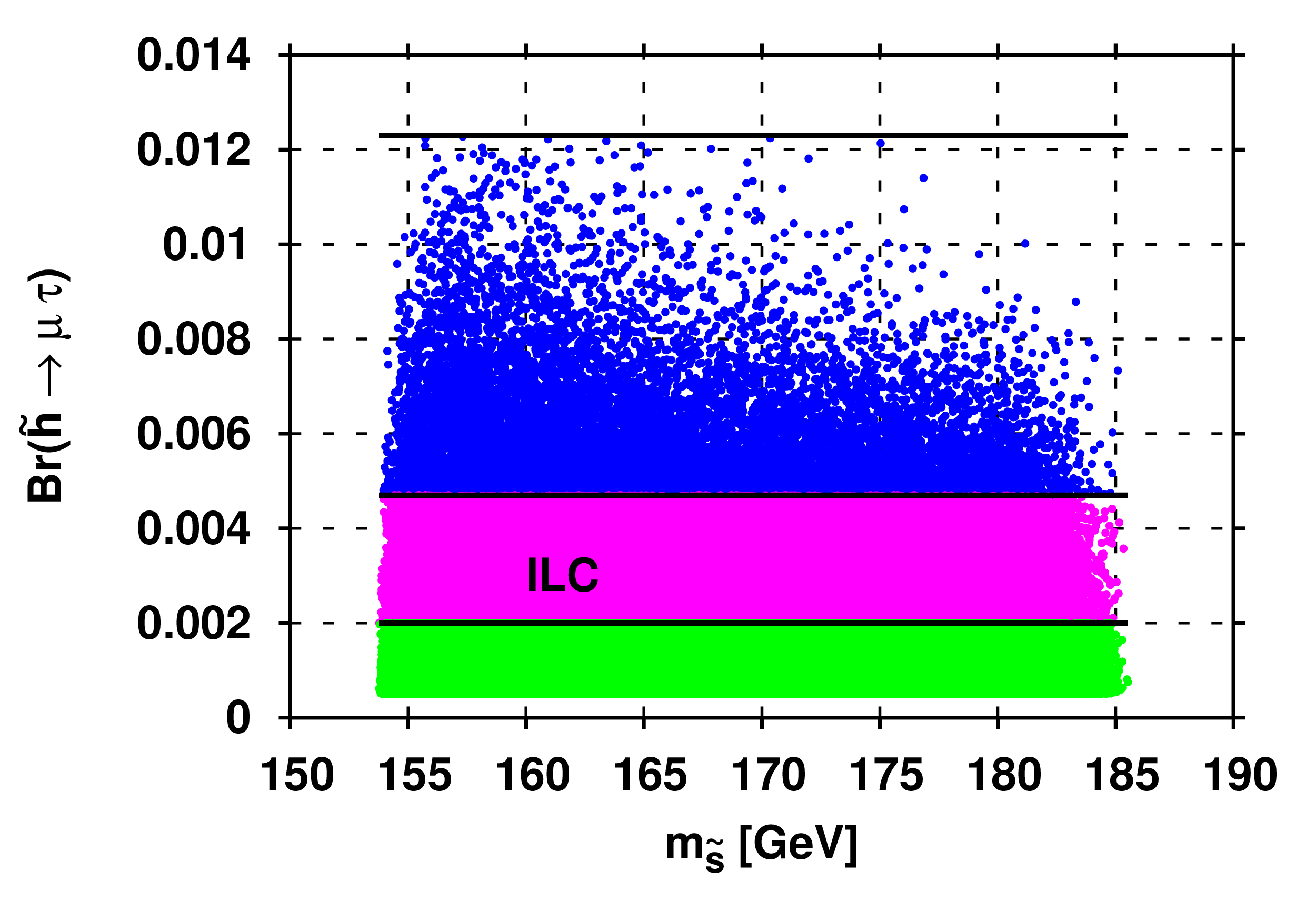}
\end{tabular}
\caption{\label{h_mutau} Scatter plots in plane $\left(m_{\tilde{s}},
  \text{Br}(\tilde{h} \to\tau\mu) \right)$ for $\sqrt{F} = 8~\text{TeV}$ and
  sgoldstino lighter (left) and heavier (right) than Higgs.} 
\end{figure}

\begin{figure}[!htb]
\begin{tabular}{cc}
\includegraphics[width=0.45\columnwidth]{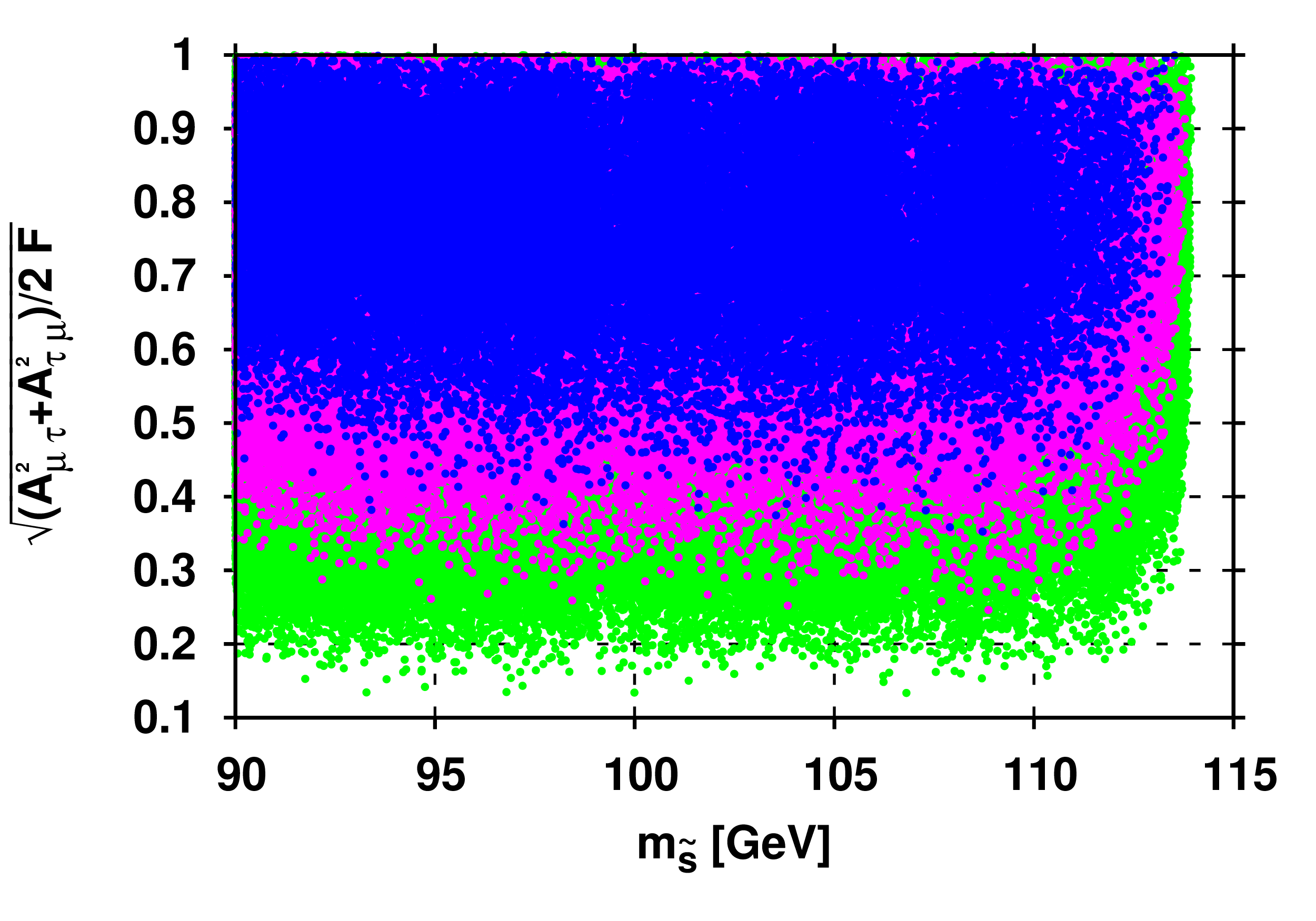}
&
\includegraphics[width=0.45\columnwidth]{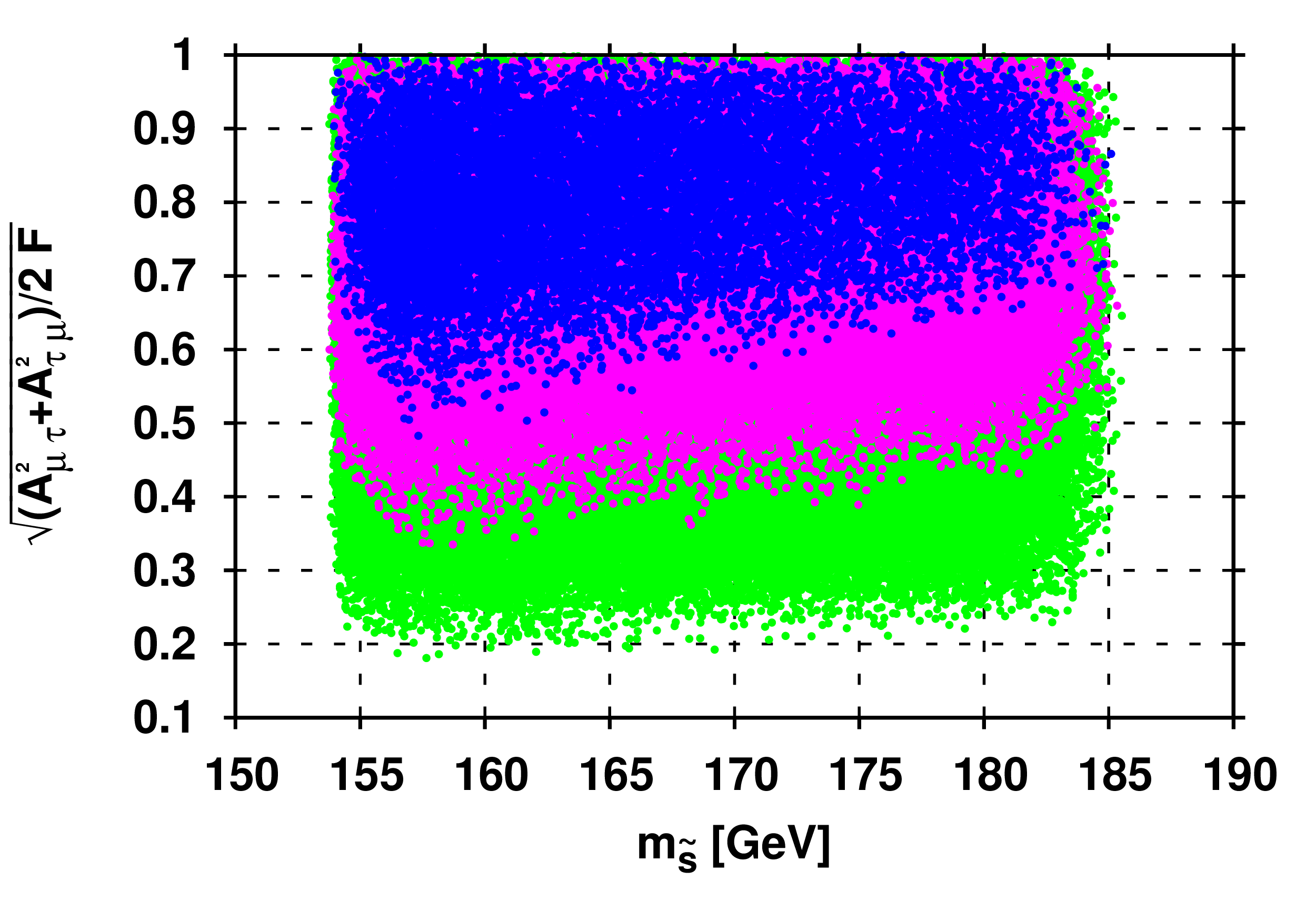}
\end{tabular}
\caption{\label{fig_pars} Scatter plots in plane $\left(m_{\tilde{s}},
  \sqrt{\frac{A_{\tau\mu}^2+A_{\mu\tau}^2}{2F}}\right)$ for $\sqrt{F}
  = 8~\text{TeV}$ and sgoldstino lighter (left) and heavier (right)
  than than the Higgs boson. By color, we show different levels of
  ${\rm Br}(\tilde{h}\to\mu\tau)$ as in
  Fig.~\ref{h_mutau}. } 
\end{figure}

In this Section we describe the results of the scan over parameter
space of the scenario with sgoldstino. In the figures below, we show
different parameters and observables for models which satisfy all
phenomenological constraints described previously. For illustrating
purposes, we present only the models with sufficiently large 
branching fraction ${\rm Br}(\tilde{h}\to\mu\tau)>5.0\cdot
10^{-4}$. By blue color we mark the models which are capable of
explaining the CMS excess, ${\rm Br}
(\tilde{h}\to\mu\tau)=8.4^{+3.9}_{-3.7}\times 10^{-3}$. In several
figures we use also purple color to mark points which lie somewhat
below the CMS excess but still have significant (more than $0.2\%$)
branching ratio. According to the latest
study~\cite{Banerjee:2016foh}, this level of branching fraction of
$\tilde{h}\to\mu\tau$ will be reachable in future experiments such as HL-LHC and ILC; see also Refs.~\cite{Mao:2015hwa,Chakraborty:2016gff}. The rest of the models
are painted in green. Corresponding predictions for ${\rm Br}(\tilde{h}\to\mu\tau)$ in the selected models are presented in  
Fig.~\ref{h_mutau} 
for light (left panel) and heavy (right panel) sgoldstino. We find a
lot of phenomenologically accepted models explaining the CMS excess. 
In Fig.~\ref{fig_pars} we present distribution of all the selected
models in $\left(m_{\tilde{s}},\sqrt{\frac{A_{\tau\mu}^2+A_{\mu\tau}^2}{2F}}\right)$-plane for lighter (left panel) and heavier (right panel) sgoldstinos. 
Sgoldstino explanation of the CMS excess requires large sgoldstino
admixture in the Higgs boson and sufficiently large 
values of the soft trilinear coupling constants $A_{\mu\tau}$ and/or
$A_{\tau\mu}$. This can present a problem for model building and we
leave this question for future study. 
Numerically, we obtain that the value of $\vert \sin{\theta} \vert$
should be larger than 0.05 (0.15) for light (heavy) sgoldstino 
for models with ${\rm Br}(\tilde{h} \to \tau \mu) > 5 \cdot
10^{-4}$. Now let us comment more on the choice of the sgoldstino mass
intervals and the value of supersymmetry breaking scale. It appears
that sgoldstino with masses larger than about 200~GeV is not capable
to explain the CMS excess in chosen parameter space (see 
Table~\ref{tbl2}). Larger sgoldstino masses result in a 
suppression of the mixing angle (see eq.~\eqref{mixing2}) and
correspondingly in a decrease of ${\rm Br}(\tilde{h} \to\mu\tau)$
below the values indicated by the CMS excess. At the same time, values
of $\sqrt{F}$ smaller than about $8$~TeV also turn out to be
disfavored by this excess and results of direct searches. Namely, at
smaller $\sqrt{F}$ the coupling constants of sgoldstino to the
SM particles increase and such sgoldstino is phenomenologically 
unacceptable.  In this case, very light sgoldstino, which decays
mostly to $b \widebar{b}$ due to large mixing with the Higgs boson,
becomes excluded by the TeVatron and LEP results. Heavier sgoldstino
with $\sqrt{F}$ smaller than about 8~TeV is excluded by the results of
the ATLAS and CMS searches for diboson resonances. If we enlarge our
parameter space by increasing, in particular, the upper bound on $\mu$
in the Table~1, we expect that somewhat lower values of $\sqrt{F}$ and
larger 
values of the sgoldstino mass will be allowed.

\begin{figure}[!htb]
\begin{tabular}{cc}
\includegraphics[width=0.45\columnwidth]{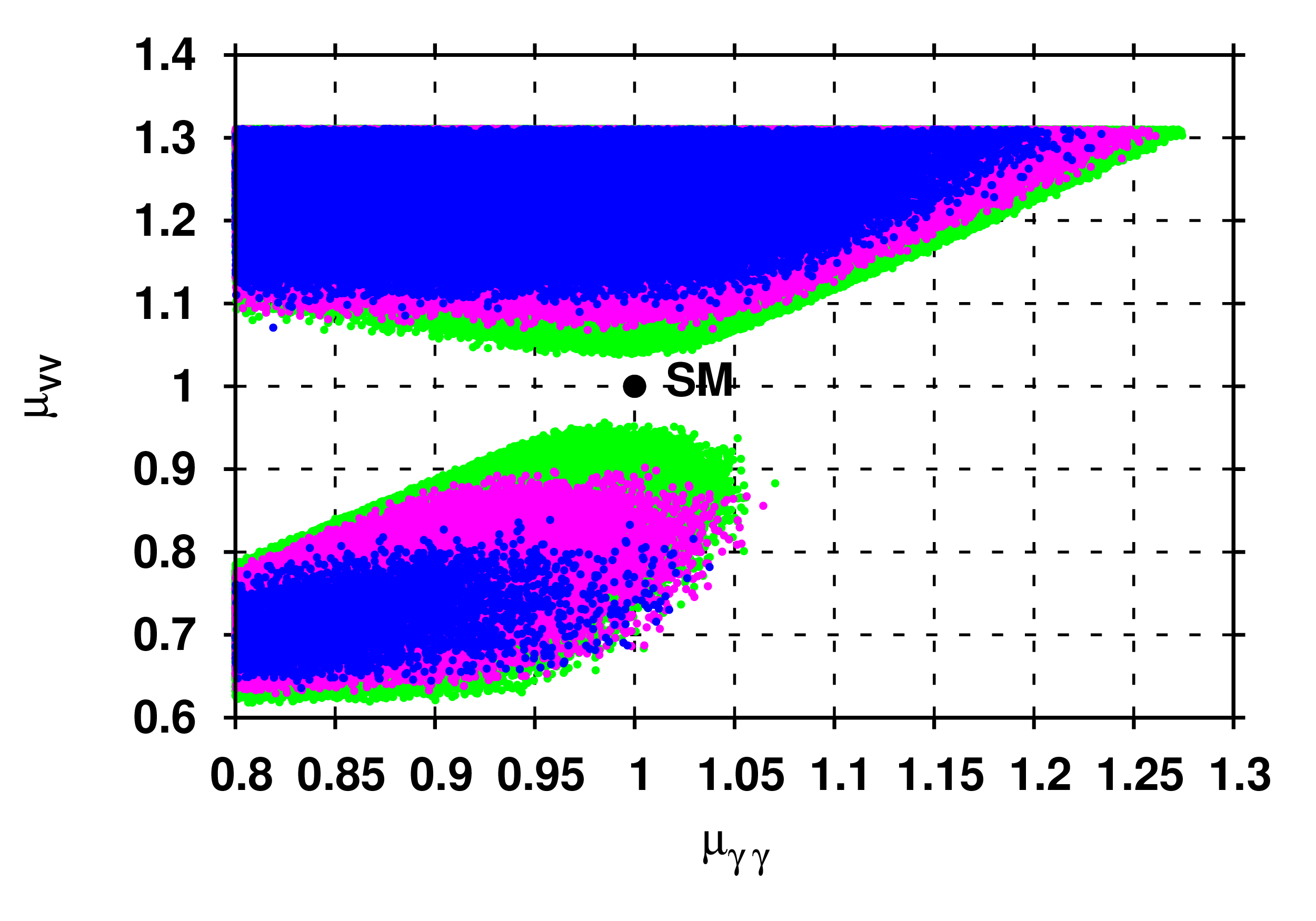}
&
\includegraphics[width=0.45\columnwidth]{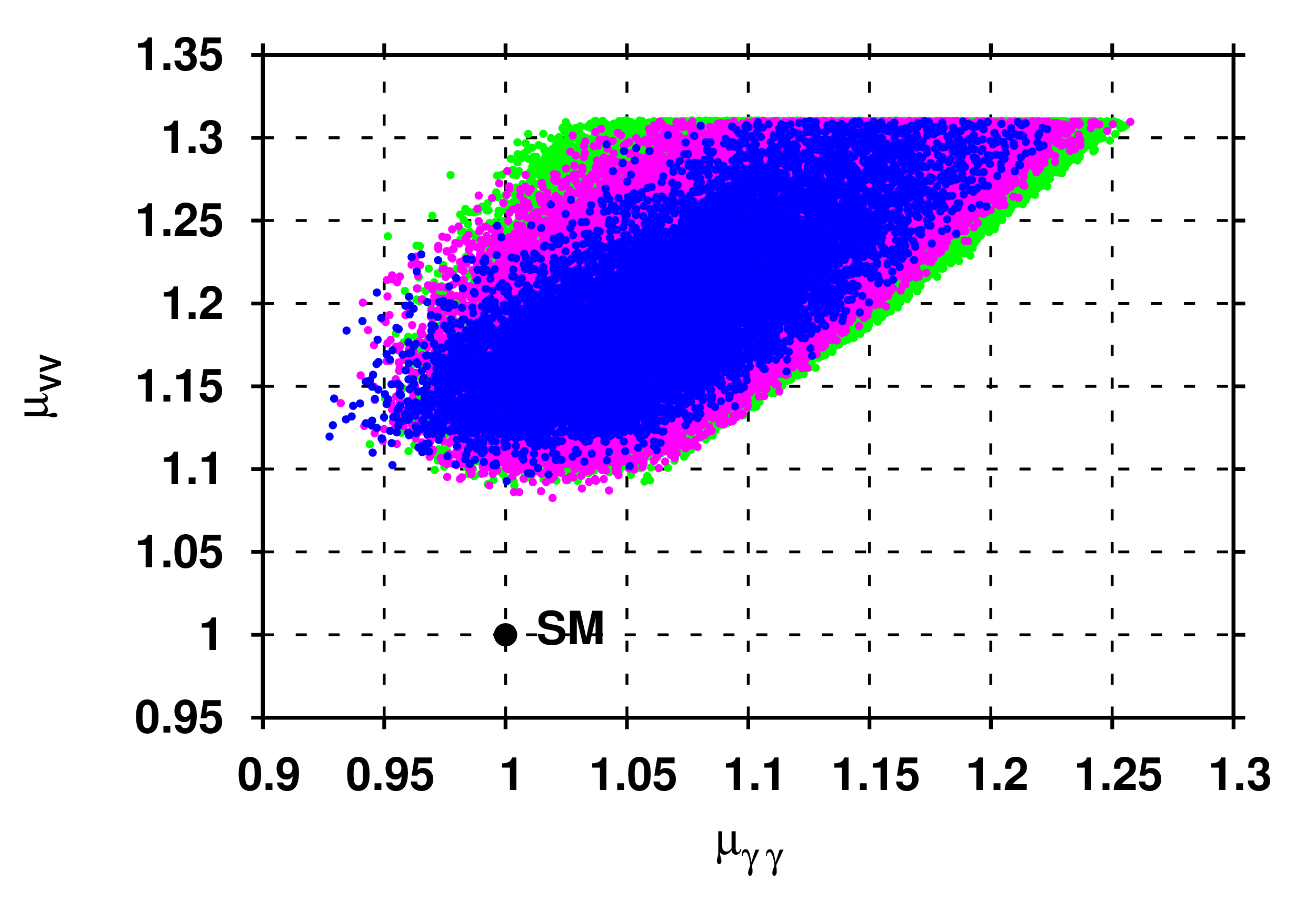}
\end{tabular}
\caption{\label{mu_VV} Scatter plots in plane $\left(\mu_{\gamma
    \gamma},\mu_{\scriptscriptstyle VV} \right)$ for $\sqrt{F} = 8~\text{TeV}$ and
  sgoldstino lighter (left) and heavier (right) than the Higgs
  boson. By color, we show different levels of
  ${\rm Br}(\tilde{h}\to\mu\tau)$ as in
  Fig.~\ref{h_mutau}.}  
\end{figure}

\noindent
In Fig.~\ref{mu_VV} 
we show the selected models in $\left(\mu_{\gamma
    \gamma},\mu_{\scriptscriptstyle VV} \right)$-plane for light (left panel) and
heavy (right panel) sgoldstino. Here $\mu_{\scriptscriptstyle VV}$ is either $\mu_{ZZ}$ or $\mu_{WW}$ (they are almost coincide for our choice of parameters). For the case of lighter sgoldstinos, two disjoint regions  correspond to the opposite 
signs of parameter $\mu$. In the case of heavier sgoldstinos, only positive values of $\mu$ are phenomenologically allowed. The deviation of $\mu_{\scriptscriptstyle VV}$  with respect to their SM values occurs mainly as a results of an increase in the Higgs-gluon coupling constant, because for the chosen parameter space couplings of sgoldstino to massive vector bosons and
$b$-quarks are smaller than those of the Higgs boson. 
The Standard Model Higgs boson interacts with massless gauge bosons
via loops only. This results in a possibility that the couplings
$g_{hgg,\rm{SM}}$ and $g_{sgg}$ can be of the same order. Estimates
show that one has $\mu_{\scriptscriptstyle VV} < 1 \, (>1)$ when
$\mu>0 \, (\mu<0)$ for $m_{\tilde{s}}<m_{\tilde{h}}$.

\begin{figure}[!htb]
\begin{tabular}{cc}
\includegraphics[width=0.45\columnwidth]{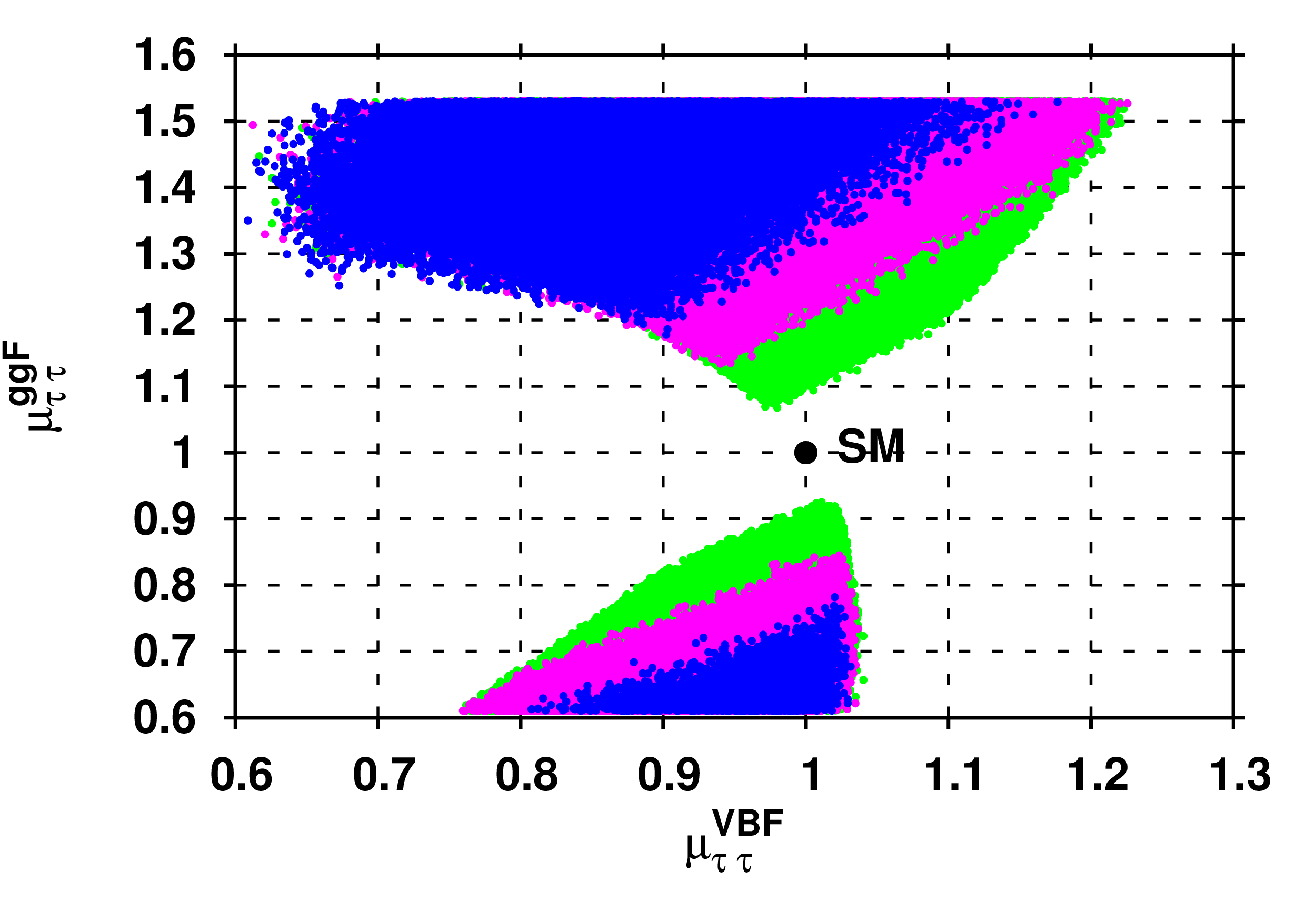}
&
\includegraphics[width=0.45\columnwidth]{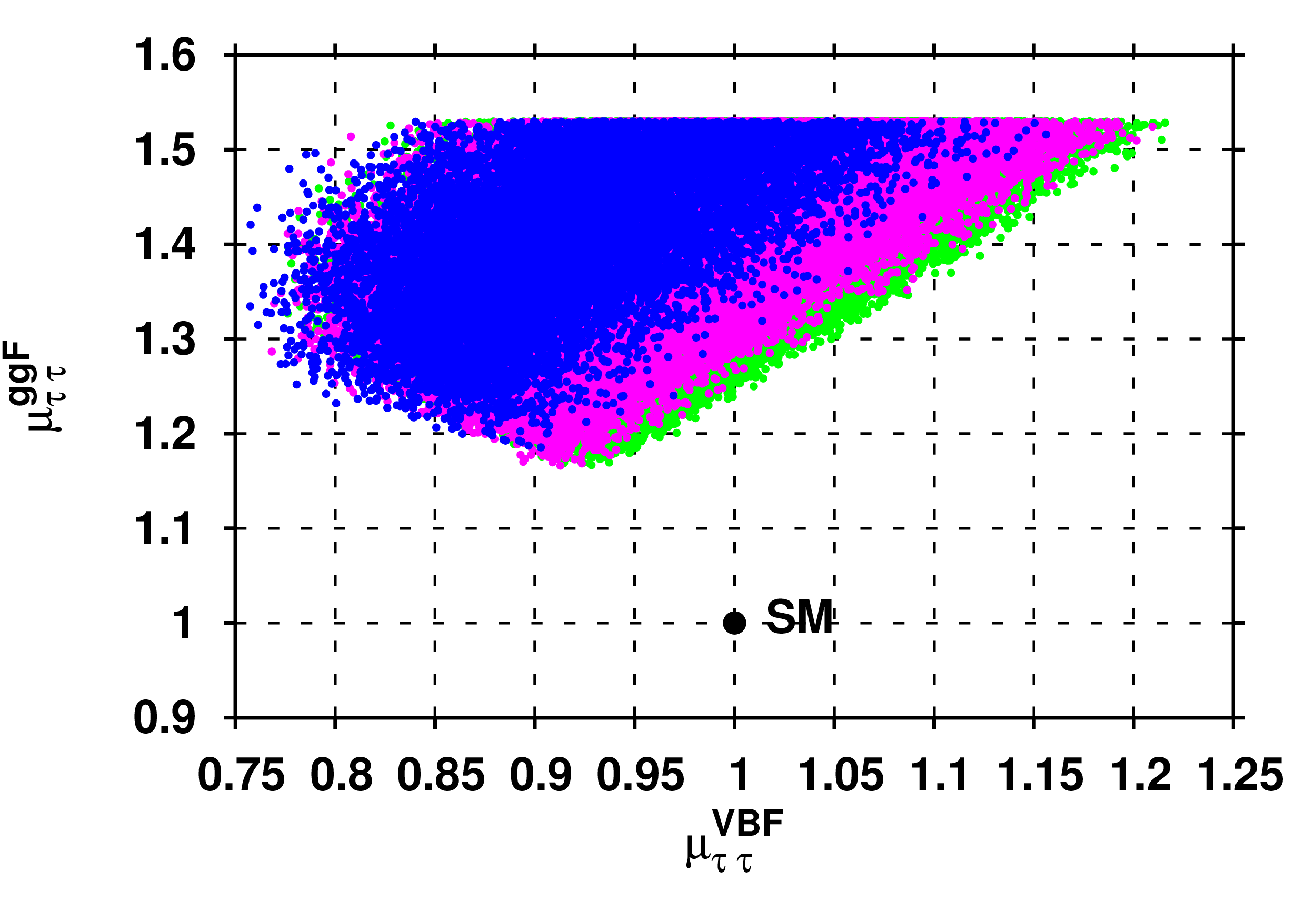}
\end{tabular}
\caption{\label{mu_fermions} Scatter plots in plane $\left(\mu_{\tau 
    \tau}^{{\rm VBF}},\mu_{\tau \tau}^{{\rm ggF}} \right)$ for
  $\sqrt{F} = 8~\text{TeV}$ and sgoldstino lighter (left) and heavier 
  (right) than the Higgs boson. By color, we show different levels of
  ${\rm Br}(\tilde{h}\to\mu\tau)$ as in
  Fig.~\ref{h_mutau}.} 
\end{figure}

\begin{figure}[!htb]
\begin{tabular}{cc}
\includegraphics[width=0.45\columnwidth]{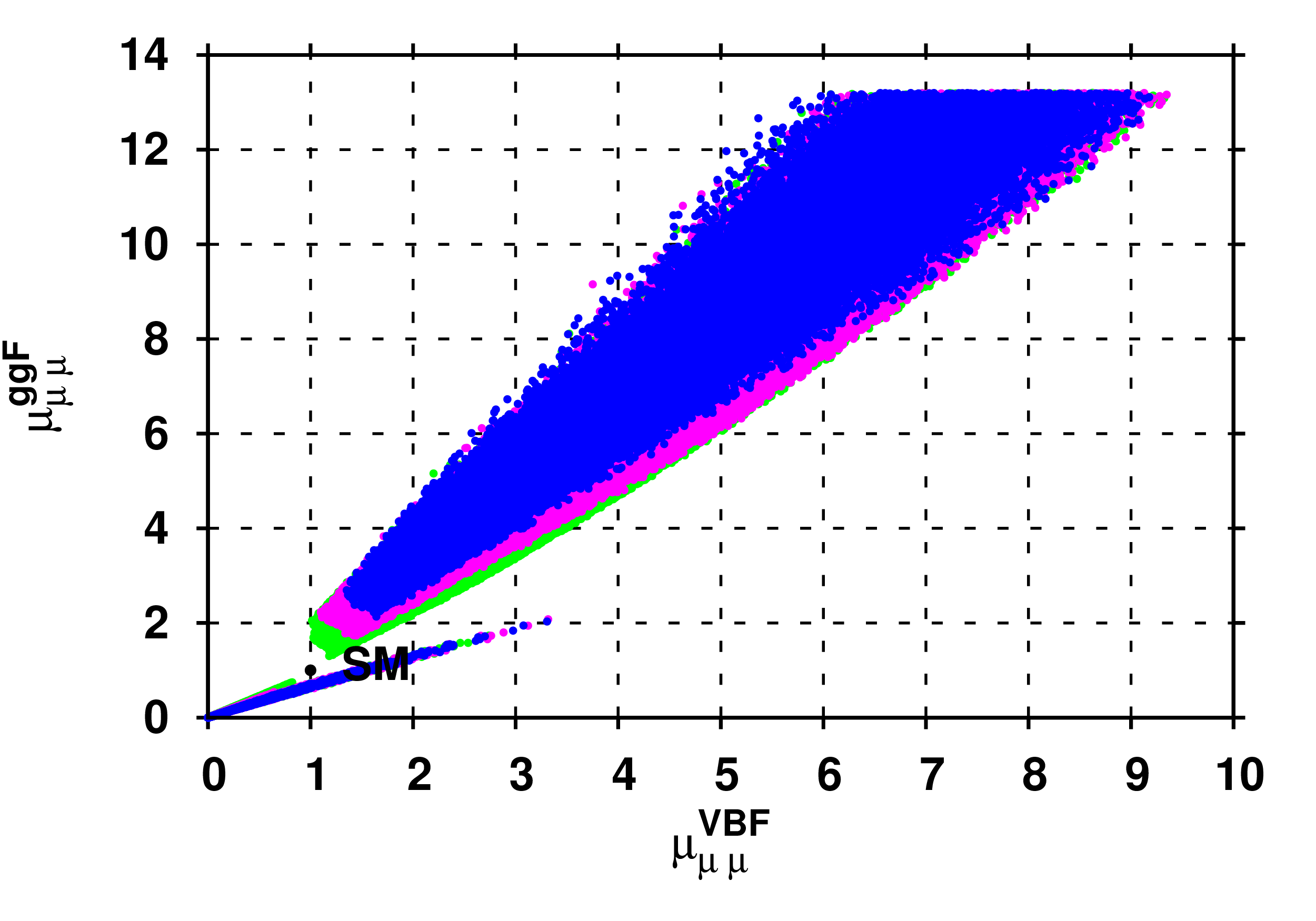}
&
\includegraphics[width=0.45\columnwidth]{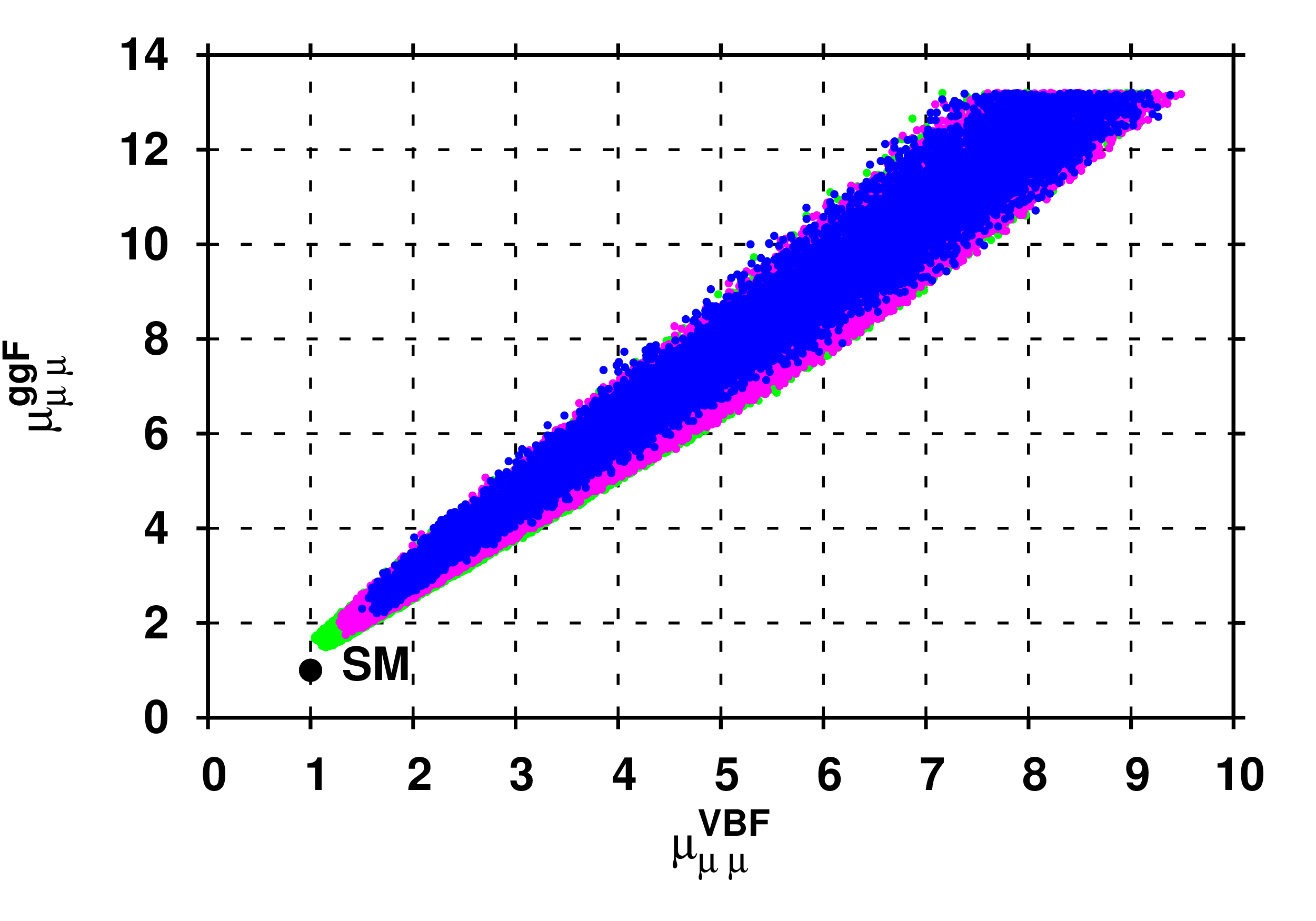}
\end{tabular}
\caption{\label{mu_mufermions} Scatter plots in plane $\left(\mu_{\mu 
    \mu}^{{\rm VBF}},\mu_{\mu \mu}^{{\rm ggF}} \right)$ for
  $\sqrt{F} = 8~\text{TeV}$ and sgoldstino lighter (left) and heavier 
  (right) than the Higgs boson. By color, we show different levels of
  ${\rm Br}(\tilde{h}\to\mu\tau)$ as in
  Fig.~\ref{h_mutau}.} 
\end{figure}

\noindent
Sgoldstino-Higgs mixing results also in changes of the signal strengths for fermionic final states. For chosen parameter space, the coupling constants of sgoldstino to $\tau$-leptons and $b$-quarks are comparable with corresponding couplings of the Higgs boson, while for muons can even exceed it. For an illustration, in
Fig.~\ref{mu_fermions} we show the scatter plot  in $\left(\mu_{\tau
  \tau}^{{\rm VBF}},\mu_{\tau \tau}^{{\rm ggF}} \right)$-plane for
$\tau^+\tau^-$ final state. Here the modifications of the signal
  strengths are mainly due to changes in the production cross section
  and the total width of the Higgs resonance. Again, disjoint regions
  for $\mu_{\tau\tau}^{\rm ggF}$ correspond to different signs of
  $\mu$.  
In figure ~\ref{mu_mufermions} 
we show the scatter plots in plane $\left(\mu_{\mu \mu}^{{\rm VBF}},\mu_{\mu \mu}^{{\rm ggF}} \right)$ for $\mu^+\mu^-$ final
state. In this case the main change is due to considerable
modification of the Higgs boson coupling to muons. For the case of
light sgoldstino and $\mu>0$, this enhancement can be partially 
compensated by a suppression of the production cross section and
corresponding models lie on the thin line in the low part of the left panel in the figure~\ref{mu_mufermions}.

\begin{figure}[!htb]
\begin{tabular}{cc}
\includegraphics[width=0.45\columnwidth]{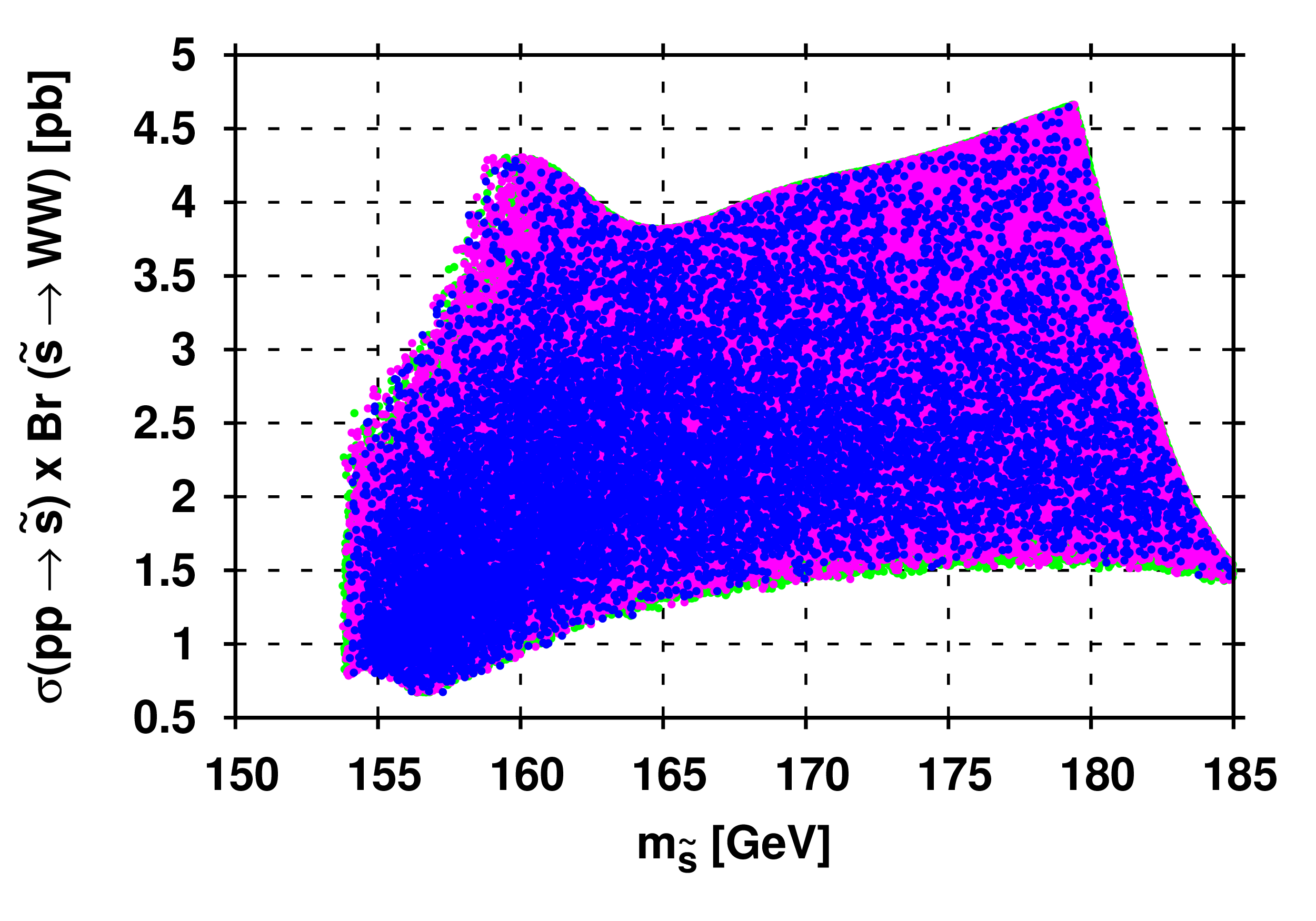}
&
\includegraphics[width=0.45\columnwidth]{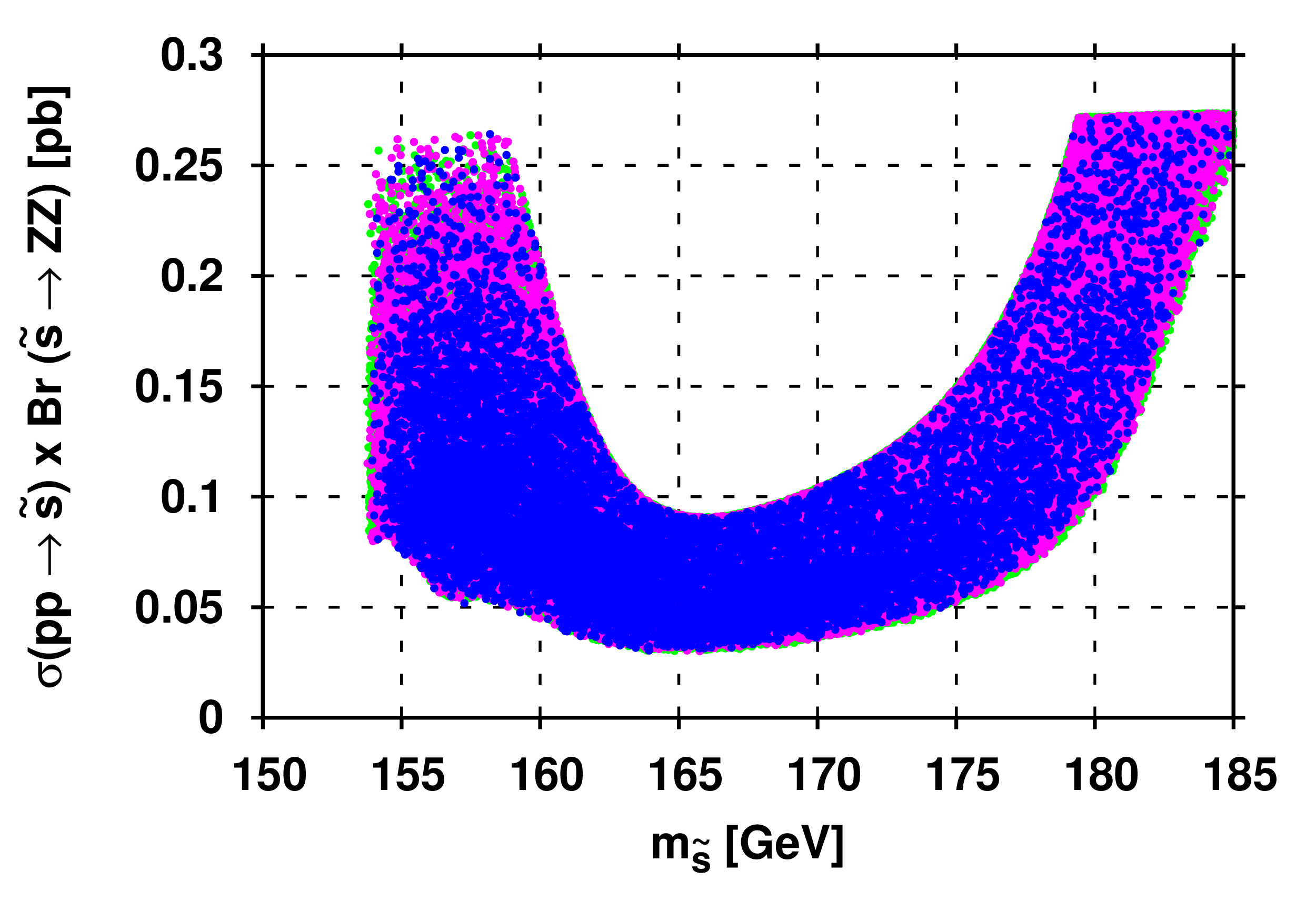}
\end{tabular}
\caption{\label{s_VV} Scatter plots in $\sigma\left(pp\to \tilde{s}\right)\times {\rm Br}\left(\tilde{s} \to W^+W^-\; {\rm or}\;ZZ\right)$-plane
for heavy sgoldstino and $\sqrt{s}=13$~TeV. By color we show different
levels of ${\rm Br}(\tilde{h}\to\mu\tau)$ as in
  Fig.~\ref{h_mutau}.}
\end{figure}

\begin{figure}[!htb]
\begin{tabular}{cc}
\includegraphics[width=0.45\columnwidth]{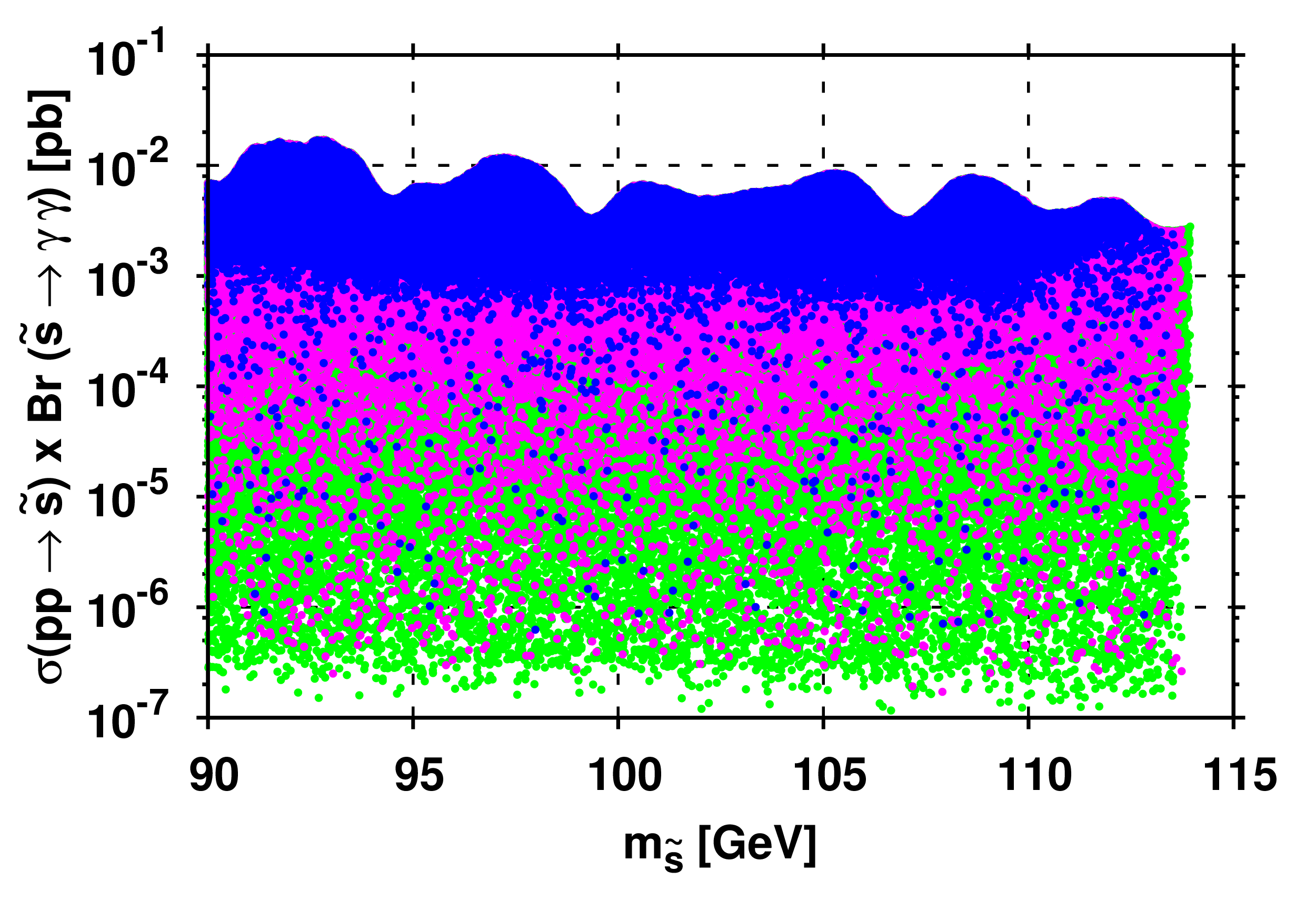}
&
\includegraphics[width=0.45\columnwidth]{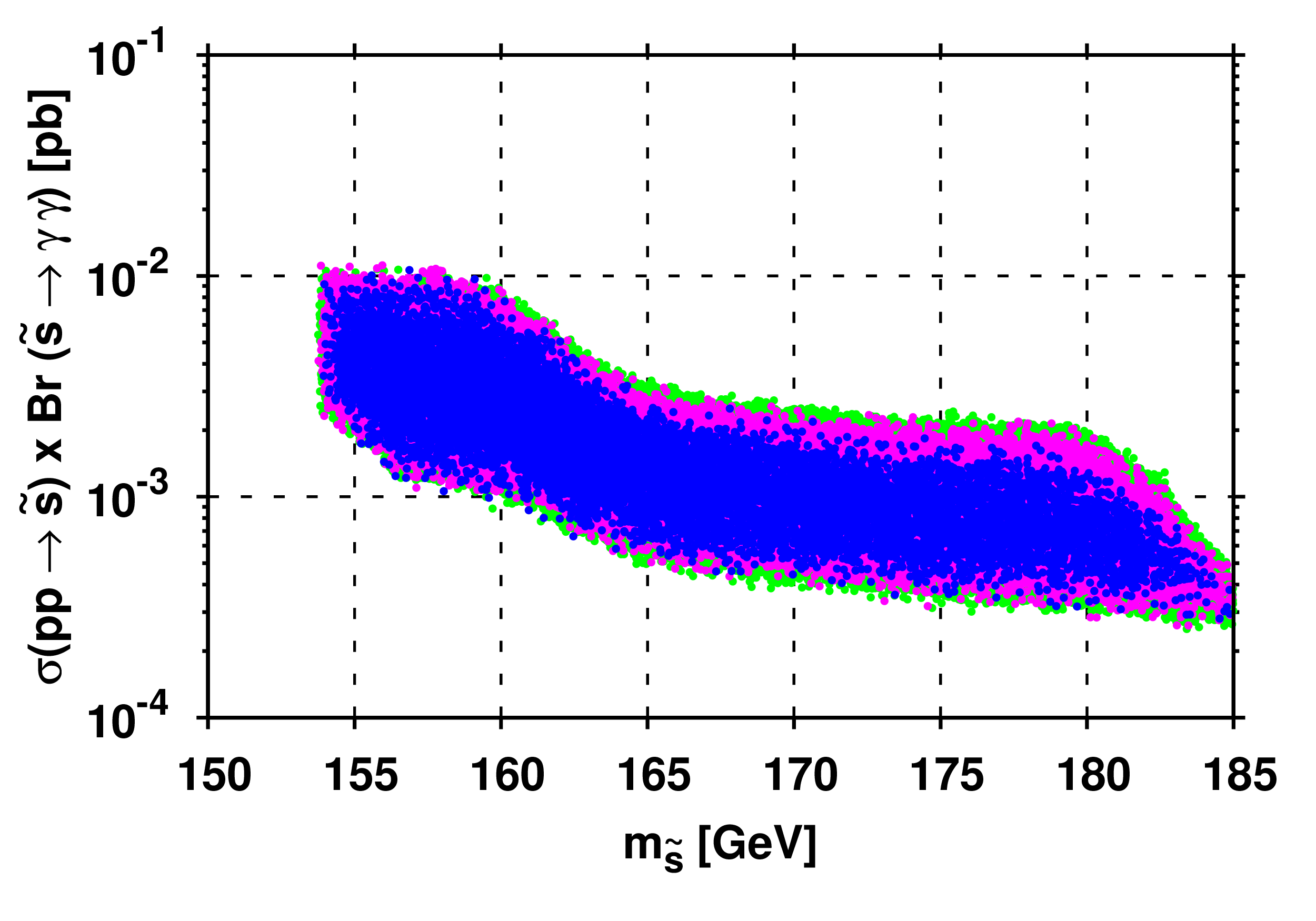}
\end{tabular}
\caption{\label{s_gmgm} Scatter plots in $\sigma\left(pp\to \tilde{s} \right)\times {\rm Br}\left(\tilde{s} \to \gamma \gamma \right)$-plane for light (left panel) and heavy (right panel) sgoldstino and $\sqrt{s}=13$~TeV. By color we show different
levels of ${\rm Br}(\tilde{h}\to\mu\tau)$ as in
  Fig.~\ref{h_mutau}.}
\end{figure}

\begin{figure}[!htb]
\begin{tabular}{cc}
\includegraphics[width=0.45\columnwidth]{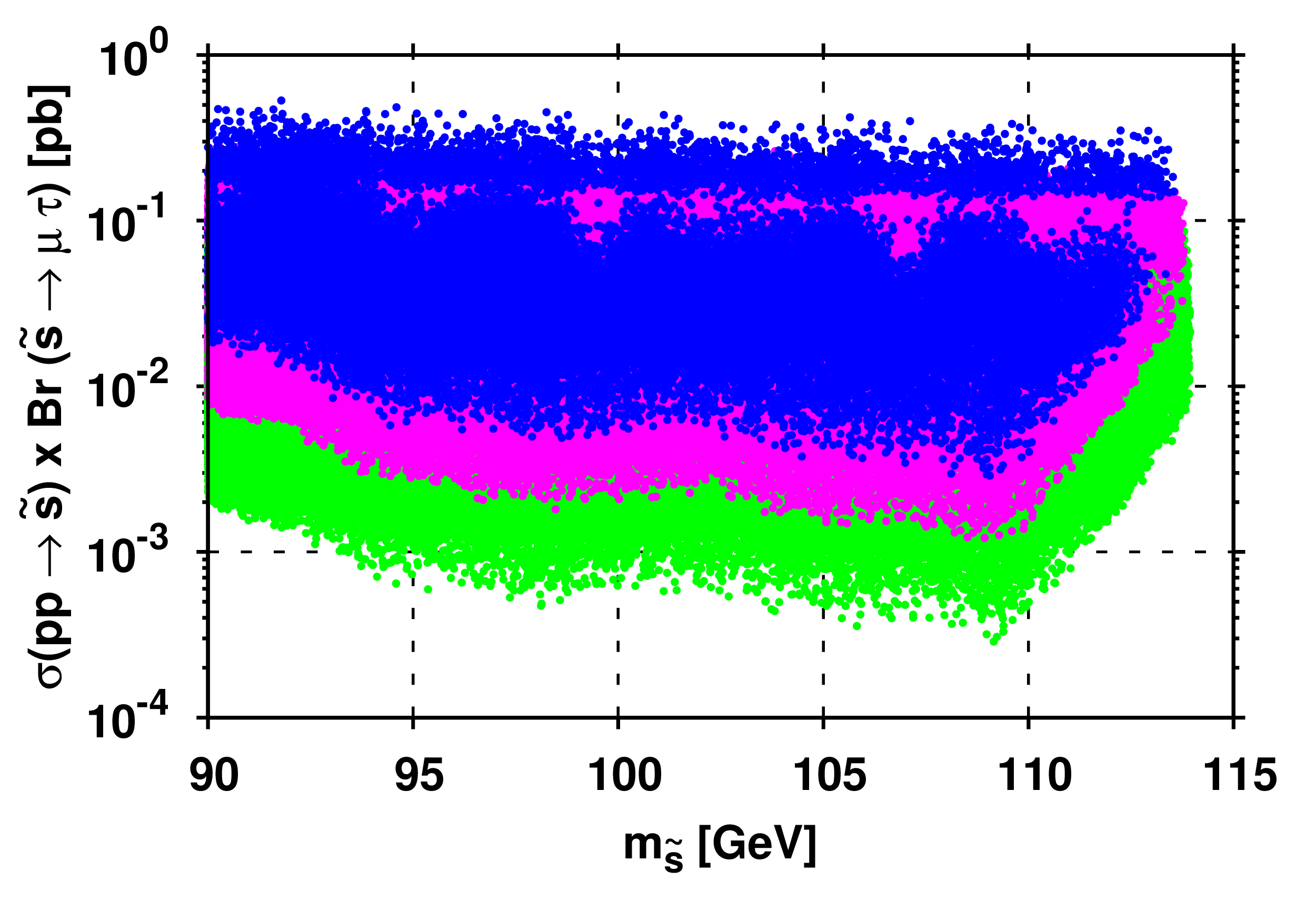}
&
\includegraphics[width=0.45\columnwidth]{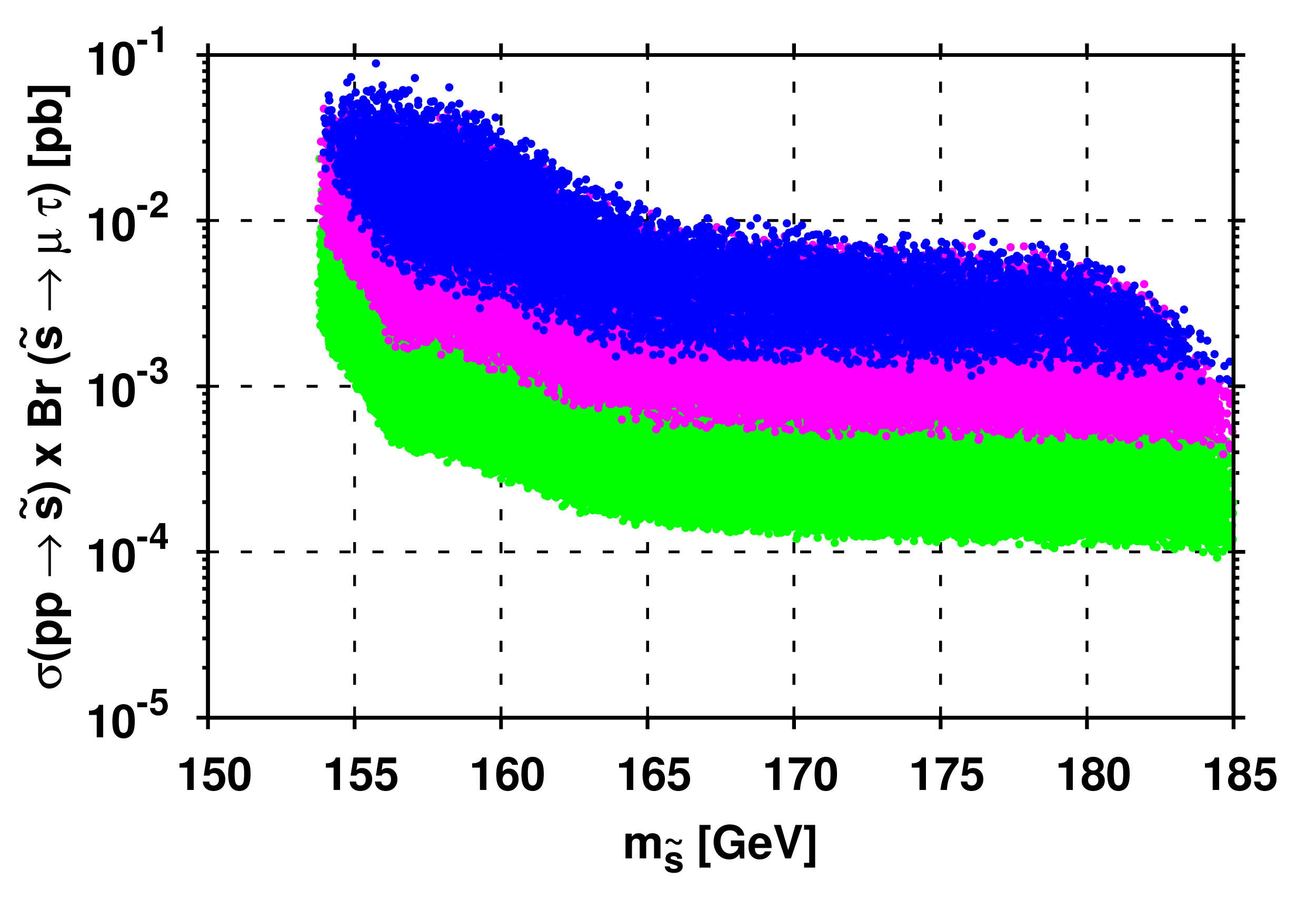}
\end{tabular}
\caption{\label{s_mutau} Scatter plots in ($m_{\tilde{s}}$, $\sigma(pp\to \tilde{s})\times {\rm Br}( \tilde{s} \to\mu\tau)$)-plane for different masses of sgoldstino and $\sqrt{s}=13$~TeV. By color we show
  different levels of ${\rm Br}(\tilde{h}\to\mu\tau)$ as in
  Fig.~\ref{h_mutau}.}
\end{figure}

\noindent
Now let us discuss the collider phenomenology of the light
sgoldstino. This scalar can reveal itself in the experiments at the
LHC as a diboson resonance. In the case of large $\sqrt{F}$ and
sufficiently large mixing of sgoldstino with the Higgs boson, the
decay pattern of  sgoldstino becomes similar to that of the Higgs
boson. It means that for heavier sgoldstino the most important decay
channels will be $W^+W^-$ and $ZZ$. Corresponding cross sections
$\sigma\left(pp\to \tilde{s}\right)\times {\rm Br}\left(\tilde{s} \to
W^+W^-\; {\rm or}\;ZZ\right)$ are presented in Fig.~\ref{s_VV}   
for the case of $\sqrt{s}=13$~TeV. Upper envelopes at these scatter
plots correspond to the current upper limits from diboson searches at the LHC. We see that predicted cross sections reach values about several picobarns for $WW$ final state and values of about 0.2~pb for $ZZ$ case which can be explored in the starting LHC run. Another important decay channel for heavier sgoldstino is decay into a pair of photons. We calculated corresponding expected cross-section for $\sqrt{s}=13$~TeV. The result is presented on the right panel of
Fig.\ref{s_gmgm}.  
Obtained values reaching 0.01--0.1~pb seem to be promising quite since they can be verified in the next run of the LHC, especially for heavy sgoldstino. Sgoldstino of lower masses with sufficiently large Higgs boson admixture decays mainly to $b\bar{b}$  but this mass region seems to be quite difficult to probe with such final state at the ATLAS and CMS experiments.
\footnote{The reason is to get rid of the overwhelming QCD background one should use here $t\bar{t}h$, vector-boson fusion or vector-boson associated production. In these cases the sgoldstino production cross section in the mass range $90 - 115~\rm{GeV}$ differs only by the factor $\sin^2 \theta$ from the corresponding production of the SM Higgs bosons with the same mass. This results in a considerable (at least an order of magnitude) signal
suppression as compared to the case of the SM Higgs boson.}
At the same time, in the considered scenario the scalar sgoldstino have large flavor-violating $\mu$--$\tau$ coupling and thus considerable branching fraction of $\tilde{s}\to\mu\tau$ decay. In Fig.~\ref{s_mutau} 
we show the cross section $\sigma\left(pp\to \tilde{s} \right)\times {\rm Br}\left( \tilde{s} \to\mu\tau\right)$ calculated at $\sqrt{s}=13$~TeV for selected models and different sgoldstino masses. We see that it reaches values about 0.1--0.2~pb for models explaining the CMS excess, which hopefully can be probed in the next runs of the LHC experiments.

\begin{figure}[!htb]
\begin{tabular}{cc}
\includegraphics[width=0.45\columnwidth]
{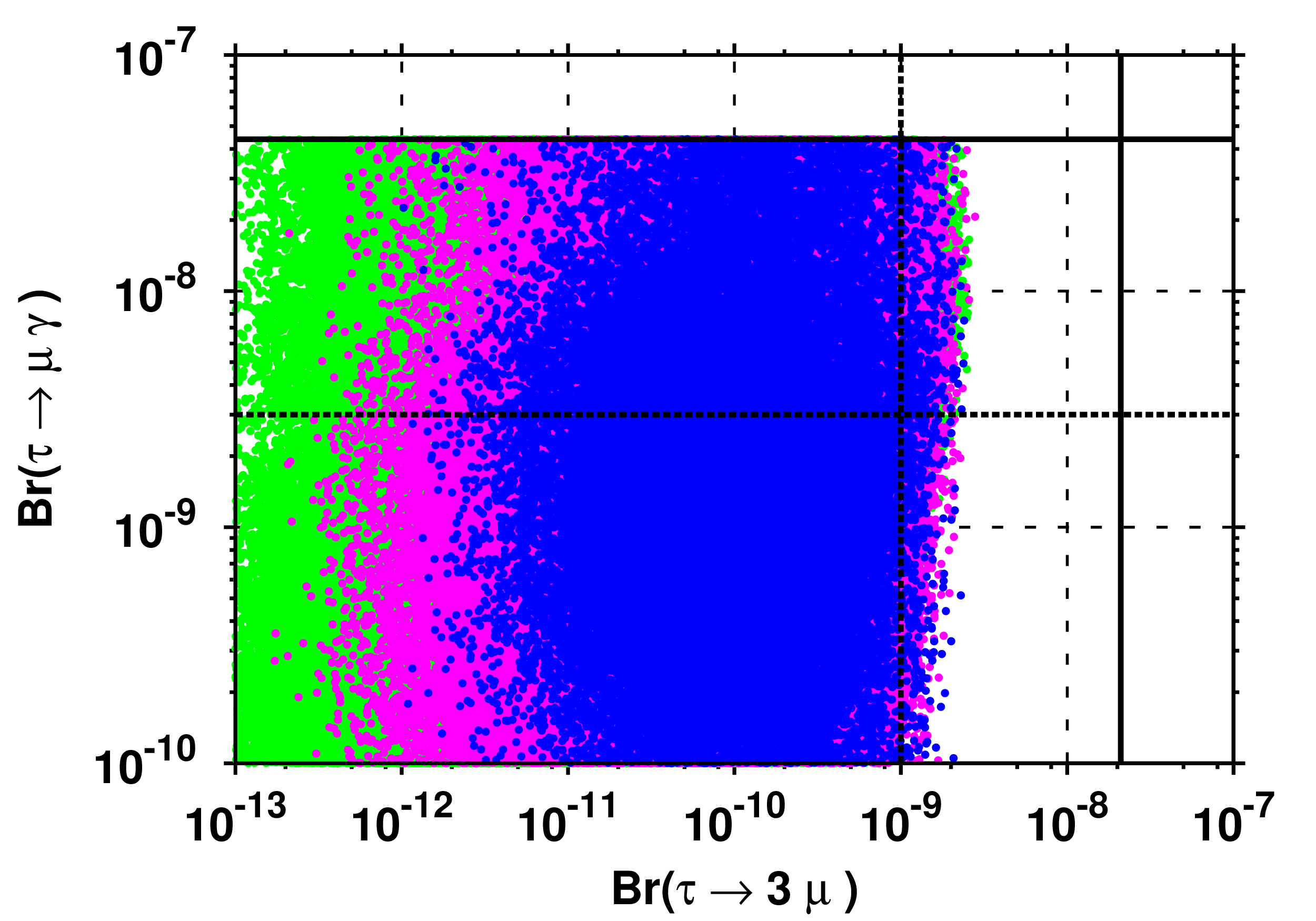}
&
\includegraphics[width=0.45\columnwidth]
{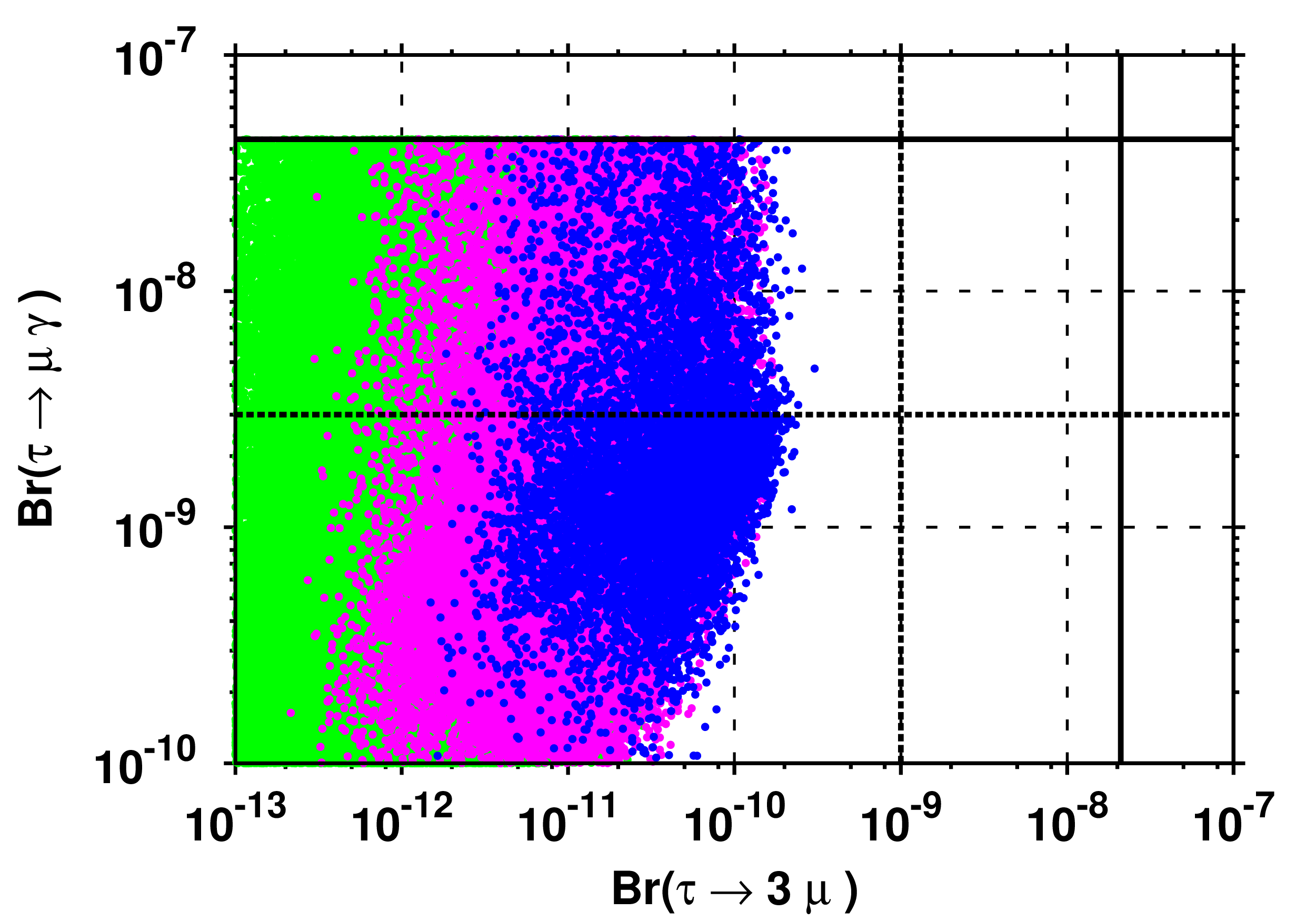}
\end{tabular}
\caption{\label{sgo_LFV} Scatter plots in plane $\left(\text{Br}(\tau
  \to \mu \gamma), \text{Br}(\tau \to 3\mu) \right)$ for $\sqrt{F} =
  8~\text{TeV}$ and sgoldstinos lighter (left) and heavier (right)
  than the Higgs boson. Solid lines correspond to present limits on
  branching fraction of both decays, whereas dashed line represents
  expected SuperKEKB sensitivity~\cite{Aushev:2010bq}. By color we
  show different levels of ${\rm Br}(\tilde{h}\to\mu\tau)$ as in
  Fig.~\ref{h_mutau}.} 
\end{figure}

\noindent
Finally, in Fig.~\ref{sgo_LFV} we present predictions for ${\rm
  Br}\left(\tau\to\mu\gamma\right)$ and ${\rm Br}\left(\tau\to
3\mu\right)$ in the model with  sgoldstino.
As we discussed in the previous section, our effective low energy
theory has limited predictive power for such observables and in the
present study they are only estimated using realistic value of the
cutoff for dominant divergent loop diagrams, $\Lambda^2=16\pi
F^2/m_{\text{sl}}^2$ (see Appendix~B and
Refs.~\cite{Brignole:2000wd,Brignole:1998uu,Brignole:1996fn}). 
The dominant contribution to ${\rm Br}(\tau\to\mu\gamma)$ comes from
the standard one-loop diagram with sfermions while for ${\rm
  Br}(\tau\to 3\mu)$ we leave only the  tree-level contribution with
sgoldstino and Higgs boson exchange. In Fig.~\ref{sgo_LFV} by solid
lines we show the current experimental bounds, while the dashed lines
show the expected SuperKEKB sensitivities to these
decays~\cite{Aushev:2010bq}. We checked that with another choice of the cutoff scale, for
  instance, $\Lambda = \sqrt{F}$ or $\Lambda = m_{sl}$, the
  predictions for the rate of $\tau\to\mu\gamma$ decay for each
  particular model in the parameter space can differ considerably. 
  But the whole picture of colored points on Fig.~\ref{sgo_LFV}
  remains almost unchanged. Thus, our analysis reveals the level of
  ${\rm Br}(\tau\to\mu\gamma)$ about $10^{-8}$ can expected in the
  considered setup.  
We see that many models with low scale
supersymmetry breaking explaining the CMS excess can be possibly
probed by these searches. However, we should stress that knowledge of
particular microscopic theory is needed to make more solid predictions
for ${\rm Br} (\tau\to\mu\gamma)$ and ${\rm Br}(\tau\to 3\mu)$ 
  for a particular point in the parameter space.

\section{Conclusions}
In this paper we showed that in models with scale of supersymmetry breaking around several TeVs and having superlight singlet goldstino and relatively light sgoldstinos (with the latter's masses around hundreds GeVs),
the Higgs boson can have considerable branching ratio of  $h\to\tau\mu$ decay. In particular, we demonstrated that the CMS excess in $h\to\tau\mu$ decay can be explained in this framework. This interesting scenario involves nonzero mixing of the lightest Higgs boson with scalar sgoldstino which can have flavor-violating couplings to SM fermions. We stress, that these features are common in the class of models in question. 

\noindent
We performed a scan over relevant parameter space of the model and
found several distinct signatures of this scenario. First of all,
due to the mixing with sgoldstino, considerable changes of the Higgs
boson signal strengths for the main search channels $\gamma\gamma$,
$ZZ$, $W^+W^-$, $\tau^+\tau^-$ and $\mu^+\mu^-$ are expected as
compared to the SM predictions. We find that for most of the models
explaining the CMS excess the signal strengths differ by more than
10\% from the SM predictions for gluon-gluon fusion production
mechanism and even more for the case of $\mu^+\mu^-$ final state.  
Also, in our setup new scalar light state, sgoldstino, with its mass
not  very far from that of the Higgs resonance is present 
in the particle spectrum. It can reveal itself in proton collisions at
the LHC decaying into the final states similar to what happens to the Higgs boson. The scalar sgoldstino can be effectively probed in searches for diboson resonances in the recently started LHC run. Predicted values of the corresponding cross sections are presented in figures~\ref{s_VV} and~\ref{s_gmgm}. Moreover, the scalar sgoldstino have also considerable flavor-violating decay with
$\mu\tau$ final state. Predictions for LFV decays $\tau\to\mu\gamma$
and $\tau\to 3\mu$ within models with low scale supersymmetry breaking are plagued from uncertainties related to precise knowledge of microscopic theory. Using some simplifying assumptions and realistic value for the cutoff of the effective theory we made an estimate for branching ratios of these decays and found obtained values to be interesting in a part of the parameter space for the nearest future experiments in this area.  

\noindent
Here we concentrated on lepton flavor violation in $\mu-\tau$ sector
motivated by the CMS results. However, we note that sgoldstino-Higgs
mixing as well as (lepton) flavour violation in sgoldstino
interactions are rather general predictions of low scale
supersymmetry. Hence, the model with light goldstino sector can result in LFV Higgs boson decay $h\to\tau e$ at similar level as predicted for $h\to\tau\mu$. 

\paragraph*{Acknowledgements.}
We thank D.~Gorbunov for useful discussions. This work is supported by 
the RSCF grant 14-22-00161. Numerical calculations have been performed
on the Computational cluster of the Theoretical division of INR RAS.

\appendix
\section{Coupling constants of $\tilde{h}$ and $\tilde{s}$. } 
In this Appendix, we present relevant expressions for the modified
coupling constants of the Higgs and sgoldstino mass states, as well as
their decay rates. In the decoupling limit of the MSSM, we are left
with the lightest Higgs boson with the following relevant effective
interactions 
\begin{equation}
\label{no22}
\mathcal{L}_{h}^{eff} = 
g_{h \gamma \gamma} \, h F_{\mu\nu} F^{\mu \nu} 
+ g_{hgg} h \, \text{tr} \, G_{\mu \nu} G^{\mu\nu}
- \frac{m_b}{\sqrt{2}v} \, h \widebar{b} b 
+ \frac{\sqrt{2} m_W^2}{v} \, h W_{\mu}^+ W^{\mu -} 
+ \frac{\sqrt{2} m_Z^2}{2v} \, h Z_{\mu} Z^{\mu}, 
\end{equation}
where $g_{h \gamma \gamma}$ and $g_{hgg}$ are the loop factors. 
Interactions between scalar sgoldstino and the SM gauge bosons and
fermions are given by
\begin{equation}
\label{no23}
\mathcal{L}_{s}^{eff} = 	
 - \frac{M_{\gamma \gamma}}{2\sqrt{2}}  \, s \, F_{\mu \nu}
F^{\mu \nu} - \frac{M_3}{2\sqrt{2}F} \, s \, \text{tr} \, G_{\mu \nu}
G^{\mu \nu} + \frac{A_{bb} \, v_d}{\sqrt{2}F} s \widebar{b} b
-\frac{M_2}{\sqrt{2}F} \, s \, W_{\mu
  \nu} W^{\mu \nu *} -\frac{M_{ZZ}}{2\sqrt{2}F} \, s \, Z_{\mu \nu}Z^{\mu \nu},
\end{equation}
where
\begin{eqnarray}
\label{no24}
\begin{aligned}
& M_{ZZ} = M_1 \, \sin^2 \theta_W + M_2 \, \cos^2 \theta_W\,, \\
& M_{\gamma \gamma} = M_1 \, \cos^2 \theta_W + M_2 \, \sin^2 \theta_W .
\end{aligned}
\end{eqnarray}
The effective interactions with photons, gluons and SM fermions have
the same form for the Higgs boson $h$ and sgoldstino $s$. As a
consequence, corresponding coupling constants for the mass state
$\tilde{h}$ will be given by the following combinations
\begin{eqnarray}
\label{effcouple}
g_{\tilde{h}\gamma \gamma} = g_{h \gamma
  \gamma,\text{SM}}^{\text{1-loop}} \, \cos \theta + \frac{M_{\gamma
    \gamma}}{2\sqrt{2}F} \, \sin \theta\,, \\
g_{\tilde{h}gg} = g_{h gg,\text{SM}}^{\text{1-loop}} \, \cos
\theta + \frac{M_{3}}{2\sqrt{2}F} \, \sin \theta 
\end{eqnarray}
for photons and gluons and
\begin{eqnarray}
\label{no26}
\begin{aligned}
& Y_{\tau \tau}^{\tilde{h}} = \frac{m_{\tau}}{\sqrt{2}v} \, \cos \theta + \frac{A_{\tau \tau}\,  v \, \cos \beta}{\sqrt{2}F} \, \sin \theta\,, \\ 
& Y_{bb}^{\tilde{h}} = \frac{m_b}{\sqrt{2}v} \, \cos \theta + \frac{A_{bb} \, v \, \cos \beta}{\sqrt{2}F} \, \sin \theta\,, \\ 
& Y_{\mu \mu}^{\tilde{h}} = \frac{m_{\mu}}{\sqrt{2}v} \, \cos \theta + \frac{A_{\mu \mu}\,  v \, \cos \beta}{\sqrt{2}F} \, \sin \theta \\ 
\end{aligned}
\end{eqnarray}
for SM fermions. The loop factors look as follows~\cite{Spira:1997dg}
\begin{eqnarray}
\begin{aligned}
\label{loops}
& g_{\text{h}\gamma \gamma,\text{SM}}^{\text{1-loop}} =
\frac{\alpha}{8\sqrt{2}\pi v} \left(A_{1}(\tau_W)+\sum_{q} N_c \,
Q_q^2 \, A_{1/2}(\tau_q)\right)\,,  \\
& g_{\text{h}gg,\text{SM}}^{\text{1-loop}} =
\frac{\alpha_s}{16\sqrt{2}\pi v} \sum_{q} A_{1/2}(\tau_q), 
\end{aligned}
\end{eqnarray}
where $A_1$ and $A_{1/2}$ are boson and fermion contributions,
respectively,
\begin{eqnarray}
\label{A1A12}
\begin{aligned}
& A_1(\tau) = -(2+3\tau+3\tau(2-\tau) f(\tau))\,, \\
& A_{1/2}(\tau) = 2\tau (1+(1-\tau) f(\tau))
\end{aligned}
\end{eqnarray}
and 
\begin{eqnarray}
\label{ftau}
\begin{aligned}
& f(\tau) = \left\{
\begin{aligned}
& \arcsin^2 \left(\frac{1}{\sqrt{\tau}}\right), & \quad \tau \geq 1 \\
& -\frac{1}{4} \left[\log
    \frac{1+\sqrt{1-\tau}}{1-\sqrt{1-\tau}}-\dot{\imath} \, \pi\right]^2, &
  \quad \tau < 1  
\end{aligned}
\right. 
\end{aligned}
\end{eqnarray}
with $\displaystyle \tau_i = \frac{4 m_i^2}{m_h^2}$.
Corresponding decay widths can be written as
\begin{align}
\label{Ghgmgm}
& \Gamma(\tilde{h}\rightarrow \gamma \gamma) = \frac{G_{\scriptscriptstyle F} \alpha^2
  m_{\tilde{h}}^3}{128\sqrt{2}\pi^3} \left\vert  \left(A_{1}(\tau_W)+\sum_{q}
N_c \, Q_q^2 \, A_{1/2}(\tau_q)\right) \cos \, \theta +
\frac{4 M_{\gamma \gamma} v \pi}{\alpha F} \sin \, \theta\right\vert^2 \,,\\
\label{Ghgg}
& \Gamma(\tilde{h}\rightarrow gg) = \frac{\alpha_s^2 \, m_{\tilde{h}}^3 \,
  G_{\scriptscriptstyle F}}{36\sqrt{2}\pi^3} \left \vert \sum_{q} \frac{3}{4}A_{1/2}(\tau_q) \, \cos
\theta + \frac{6 M_3 \pi v}{\alpha_s \, F} \, \sin \theta \right
\vert^2\,, \\
\label{hff}
& \Gamma(\tilde{h} \rightarrow \widebar{f}f) = \frac{m_{\tilde{h}} \, (Y_{ff}^{\tilde{h}})^2}{8 \pi} \left(1-
\frac{4m_f^2}{m_{\tilde{h}}^2}\right)^{\frac{3}{2}}\,.
\end{align}
The case of interaction with $W$ and $Z$ bosons is more involved,
because of different type of operators in
Eqs.~\eqref{no22},\eqref{no23}.  Corresponding couplings for
$\tilde{h}$ can be conveniently written in the momentum space as
follows 
\begin{equation}
\label{tildehVV}
g_{\tilde{h}VV}^{\mu \nu} = g_{hVV}^{\mu \nu} \cos \theta +
\frac{M_{VV}}{\sqrt{2}F}\left((k_{V_1},k_{V_2})\eta^{\mu \nu}-k^{V_1 \mu}
k^{V_2 \nu}\right) \sin  \theta, 
\end{equation}
where
\begin{equation}
\label{hVV}
g_{hVV}^{\mu \nu} = \frac{2 m_V^2}{v} \eta^{\mu \nu}
\end{equation}
and $M_{VV}$ is either $M_{2}$ or $M_{ZZ}$ for $W$ and $Z$ bosons,
respectively. The expression for the decay width~\cite{Romao:1998sr} which 
takes into account possibility of the virtual massive vector boson
production looks as
\begin{eqnarray}
\label{WidthG}
\begin{aligned}
\Gamma (\tilde{h} \rightarrow V^* V^* \rightarrow \text{leptons}) =
\frac{1}{\pi} & \int_{0}^{m_{\tilde{h}}^2} \, d \Delta_i^2 \frac{\Gamma_V M_V}{\vert D(\Delta_i^2)\vert^2} \frac{1}{\pi} \\
& \times \int_{0}^{\left(m_{\tilde{h}}-\sqrt{\Delta_i^2}\right)^2} \, d \Delta_j^2 \frac{\Gamma_V M_V}{\vert D(\Delta_j^2) \vert^2} \, \Gamma_0^V (\Delta_i,\Delta_j,m_{\tilde{h}},\theta), 
\end{aligned}
\end{eqnarray}
where
\begin{eqnarray}
\begin{aligned}
  \nonumber
  \Gamma_0^V (\Delta_i, \Delta_j, m_{\tilde{h}}, \theta) = \delta_V \frac{G_{\scriptscriptstyle} m_{\tilde{h}}^3}{16 \pi \sqrt{2}} \sqrt{\lambda(\Delta_i^2, \Delta_j^2, m_{\tilde{h}}^2)}  \Biggl[ & \cos^2 \theta \left ( \lambda(\Delta_i, \Delta_j, m_{\tilde{h}})+12 \frac{\Delta_i^2 \Delta_j^2}{m_{\tilde{h}}^4} \right ) \\
    \nonumber
  & + X(\Delta_i^2, \Delta_j^2, \theta) \Biggr ],
\end{aligned}
\end{eqnarray}
\begin{equation}
  \nonumber
D(\Delta^2) = \Delta^2 - m_V^2 + \dot{\imath} m_V \, \Gamma_V,
\end{equation}
\begin{equation}
 \label{WidthG0}
 \begin{aligned}
   & \lambda(\Delta_i^2, \Delta_j^2, m_{\tilde{h}}^2) =
  \left ( 1-\frac{\Delta_i^2}{m_{\tilde{h}}^2}-\frac{\Delta_j^2}{m_{\tilde{h}}^2} \right )^2 -  4 \frac{\Delta_i^2 \Delta_j^2}{m_{\tilde{h}}^4}\,, 
    \end{aligned}
\end{equation}
\begin{eqnarray}
  \begin{aligned}
    \nonumber
X(\Delta_i^2, \Delta_j^2, \theta) = \frac{\Delta_i^2
    \Delta_j^2}{m_{\tilde{h}}^4} \,  \Omega \, \sin \theta \Biggl[& 12 \cos \theta
      \left (-\Delta_i^2 - \Delta_j^2 + m_{\tilde{h}}^2 \right )+ \\
&      4~\Omega ~\sin \theta \left(\frac{\left(\Delta_i^2+\Delta_j^2-m_{\tilde{h}}^2 \right)^2}{2}+\Delta_i^2 \, \Delta_j^2 \right) \Biggr ]
\end{aligned}
\end{eqnarray}
and $ \Omega =\displaystyle \frac{M_{\scriptscriptstyle{VV}} v}{F}$.
In formulas (\ref{WidthG0}), $\delta=1(2)$ for $Z(W)$ bosons and
$\Delta_{i,j}$ is a four-momentum of off-shell particles $V^*$. 

\noindent
The same expressions \eqref{WidthG} and \eqref{WidthG0} for decay
widths are applied for sgoldstinos with the substitutions $\cos
\theta \rightarrow \sin \theta$ and $\sin \theta \rightarrow -\cos
\theta$. 



\section{Contributions to $\tau\to\mu\gamma$ decay}
In this Appendix, we present expressions for different contributions
to the Wilson coefficients $c_L$ and $c_R$ in Eq \eqref{Lefftmg} which
we use to estimate the branching ratio of $\tau\to\mu\gamma$ decay. We
start with the standard SUSY part arising from slepton sector (see Fig
\ref{fig:lsusy}) which numerically gives dominant contribution for
almost all models selected in our scan. The $6 \times 6$ slepton
squared mass matrix in electroweak interaction basis ($\tilde{e}_L, \, 
\tilde{\mu}_L, \, \tilde{\tau}_L, \, \tilde{e}_R, \, \tilde{\mu}_R, \,
\tilde{\tau}_R$) can be written in terms of 3 non-diagonal $3 \times
3$ matrices \cite{Arana-Catania:2013ggc} \footnote{For clarity, we
  replace letters denoting generations  with corresponding 
  numbers $\text{e} \rightarrow 1, \, \mu \rightarrow 2, \, \tau
  \rightarrow 3$.  }
\begin{equation}
\label{no17}
M_{\tilde{l}}^2 = 
\left(
\begin{matrix}
M_{\tilde{l}LL}^2 & M_{\tilde{l}LR}^2 \\
M_{\tilde{l}RL}^{2 \dagger} & M_{\tilde{l}RR}^2 \\
\end{matrix}
\right),
\end{equation}
where
\begin{eqnarray}
\label{no18}
\begin{aligned}
& M_{\tilde{l}LL ij}^2 = m_{\tilde{L}ij}^2 + \left(m_{l_i}^2 +
  \left(\sin^2 \theta_{W} - \frac{1}{2}\right)M_Z^2 \, \cos 2 \beta
  \right) \, \delta_{ij},  \\ 
& M_{\tilde{l}RR ij}^2 = m_{\tilde{E}ij}^2 + (m_{l_i}^2 - \sin^2
  \theta_{W} \, M_Z^2 \, \cos 2 \beta) \, \delta_{ij}, \\ 
& M_{\tilde{l}LR ij}^2 = v_{d} \, A_{ij} - m_{l_i} \,\mu \, \tan
  \beta \, \delta_{ij}.
\end{aligned}
\end{eqnarray}
The matrices $m_{\tilde{L}}$, $m_{\tilde{E}}$, $A$  can be
parametrized as 
\begin{eqnarray}
\label{no19}
\begin{aligned}
& m_{\tilde{L}}^2 = \left(
\begin{matrix}
m_{\tilde{L}_1}^2 & \delta_{12}^{LL} \, m_{\tilde{L}_1}
m_{\tilde{L}_2} & \delta_{13}^{LL} \, m_{\tilde{L}_1} m_{\tilde{L}_3}
\\ 
\delta_{21}^{LL} m_{\tilde{L}_2} m_{\tilde{L}_1} & m_{\tilde{L}_2}^2 &
\delta_{23}^{LL} \, m_{\tilde{L}_2} m_{\tilde{L}_3} \\ 
\delta_{31}^{LL} m_{\tilde{L}_3} m_{\tilde{L}_1} & \delta_{32}^{LL} \,
m_{\tilde{L}_3} m_{\tilde{L}_2} & m_{\tilde{L}_3}^2 \\ 
\end{matrix}
\right),
\\
\\
& v_{d} \, A = \left(
\begin{matrix}
m_e \, A_{ee} & \delta_{12}^{LR} \, m_{\tilde{L}_1}
m_{\tilde{E}_2} & \delta_{13}^{LR} \, m_{\tilde{L}_1} m_{\tilde{E}_3}
\\ 
\delta_{21}^{LR} m_{\tilde{L}_2} m_{\tilde{E}_1} & m_{\mu} \,
A_{\mu \mu} & \delta_{23}^{LR} \, m_{\tilde{L}_2} m_{\tilde{E}_3}
\\ 
\delta_{31}^{LR} m_{\tilde{L}_3} m_{\tilde{E}_1} & \delta_{32}^{LR} \,
m_{\tilde{L}_3} m_{\tilde{E}_2} & m_{\tau} \, A_{\tau \tau} \\ 
\end{matrix}
\right),
\\
\\
& m_{\tilde{E}}^2 = \left(
\begin{matrix}
m_{\tilde{E}_1}^2 & \delta_{12}^{RR} \, m_{\tilde{E}_1}
m_{\tilde{E}_2} & \delta_{13}^{RR} \, m_{\tilde{E}_1} m_{\tilde{E}_3}
\\ 
\delta_{21}^{RR} m_{\tilde{E}_2} m_{\tilde{E}_1} & m_{\tilde{E}_2}^2 &
\delta_{23}^{RR} \, m_{\tilde{E}_2} m_{\tilde{E}_3} \\ 
\delta_{31}^{RR} m_{\tilde{E}_3} m_{\tilde{E}_1} & \delta_{32}^{RR} \,
m_{\tilde{E}_3} m_{\tilde{E}_2} & m_{\tilde{E}_3}^2 \\ 
\end{matrix}
\right).
\end{aligned}
\end{eqnarray}
We assume that the  LFV contribution comes from the trilinear
couplings $A_{\mu\tau}$ and $A_{\tau\mu}$ only. Hence, we take 
$\delta_{ij}^{LL}=\delta_{ij}^{RR}=0$ for $i\ne j$ and also
$\delta_{12}^{LR}=\delta_{21}^{LR}=\delta_{13}^{LR}=\delta_{31}^{LR}=0$.  
For simplicity, we assume a common mass scale $m_{\text{sl}}$ for
  right and left sleptons for all generations. Then we use general
expression for contributions from SUSY particles
obtained~\cite{Arana-Catania:2013ggc,Paradisi:2005fk} in Mass Insertion Approximation 
\begin{eqnarray}
\label{1loopsusy}
\begin{aligned}
& c_{L}^{\text{SUSY}}= \frac{5\pi}{3} \frac{\alpha_2}{c_{\text{w}}^2}
  v \,  A_{\tau \mu} \, \cos \beta \, \frac{M_1}{m_{\tau}}
  \frac{1}{m_{\tilde{R}}^2-m_{\tilde{L}}^2}
  \left(\frac{f_{3n}(a_L)}{m_{\tilde{L}}^2}-\frac{f_{3n}(a_R)}{m_{\tilde{R}}^2}\right)
  + \scriptstyle(\text{``LL contribution"}), \\ 
& c_{R}^{\text{SUSY}}= \frac{5\pi}{3} \frac{\alpha_2}{c_{\text{w}}^2}
  v \,  A_{\mu \tau} \, \cos \beta \, \frac{M_1}{m_{\tau}}
  \frac{1}{m_{\tilde{R}}^2-m_{\tilde{L}}^2}
  \left(\frac{f_{3n}(a_L)}{m_{\tilde{L}}^2}-\frac{f_{3n}(a_R)}{m_{\tilde{R}}^2}\right)
  + \scriptstyle(\text{``RR contribution"}) .
\end{aligned}
\end{eqnarray}
In this expression, $m_{\tilde{L}}$ and $m_{\tilde{R}}$ are the average
slepton masses in ``left" and ``right" sectors, respectively,
$a_{\scriptstyle L,R} = \displaystyle
\frac{M_1^2}{m_{\tilde{L},\tilde{R}}^2}$ and $f_{3n}$ is the loop 
function from neutralino contribution \cite{Arana-Catania:2013ggc,Paradisi:2005fk}  
\begin{equation}
\label{f3n}
f_{3n}(a) = \frac{1+2a \, \log a-a^2}{2(1-a)^3} .
\end{equation}
Using the simplifying assumptions discussed above and taking the limit  
$m_{\tilde{L}}^2 - m_{\tilde{R}}^2 \rightarrow 0$, the expression
reduces to  
\begin{equation}
\label{expres}
\frac{1}{m_{\tilde{R}}^2-m_{\tilde{L}}^2}
\left(\frac{f_{3n}(a_L)}{m_{\tilde{L}}^2}-\frac{f_{3n}(a_R)}{m_{\tilde{R}}^2}\right)
\longrightarrow \frac{2 f_{2n}(a)}{m_{\text{sl}}^4}, 
\end{equation}
where $f_{2n}(a)$ is another neutralino loop function \cite{Arana-Catania:2013ggc}
\begin{equation}
\label{f2n}
f_{2n}(a) = \frac{-5a^2+4a+1+2a(a+2)\log a}{4(1-a)^4}\,.
\end{equation}

\noindent
Now let us describe contributions from the diagrams with the Higgs
boson and sgoldstino presented in Fig.~\ref{fig:1lh}. The leading
order terms in the expansion in powers of $m_{\mu}/m_{\tilde{h}}$
($m_{\mu}/m_{\tilde{s}}$) and $m_{\tau}/m_{\tilde{h}}$ ($m_{\tau}/m_{\tilde{s}}$) for the
diagrams with internal Higgs-like state $\tilde{h}$ 
\cite{Harnik:2012pb}  are
\begin{eqnarray}
\label{C1looph}
\begin{aligned}
& c_{L}^{\text{1-loop},\tilde{h}} \simeq \frac{1}{12 \,
    m_{\tilde{h}}^2} Y_{\tau \tau}^{\tilde{h}} \, Y_{\tau
    \mu}^{\tilde{h}} \left(-4 + 3\log
  \frac{m_{\tilde{h}}^2}{m_\tau^2}\right) + \frac{1}{12 \,
    m_{\tilde{h}}^2} Y_{\mu \mu}^{\tilde{h}} \, Y_{\tau
    \mu}^{\tilde{h}} \left(-4 + 3\log 
  \frac{m_{\tilde{h}}^2}{m_\mu^2}\right)\,, \\ 
& c_{R}^{\text{1-loop},\tilde{h}} \simeq \frac{1}{12 \, m_{\tilde{h}}^2} Y_{\tau
    \tau}^{\tilde{h}} \, Y_{\mu \tau}^{\tilde{h}} \left(-4 + 3\log
  \frac{m_{\tilde{h}}^2}{m_\tau^2}\right) +  \frac{1}{12 \,
    m_{\tilde{h}}^2} Y_{\mu \mu}^{\tilde{h}} \, Y_{\mu
    \tau}^{\tilde{h}} \left(-4 + 3\log
  \frac{m_{\tilde{h}}^2}{m_\mu^2}\right) \\  
\end{aligned}
\end{eqnarray}
and similar expressions with replacements $m_{\tilde{h}} \to
m_{\tilde{s}}$ and $Y_{ab}^{\tilde{h}}\to Y_{ab}^{\tilde{s}}$ are hold
for the case of intermediate sgoldstino, i.e. for
$c_{L}^{1-loop,\tilde{s}}$ and 
$c_{R}^{1-loop,\tilde{s}}$.  
Let us note, that sgoldstino and Higgs boson contributions can be of the same magnitude: on the one hand the Higgs boson contribution is  enhanced by a factor of $\sim \left(m_{\tilde{s}}/m_{\tilde{h}}\right)^2$ but on the other
hand it is suppressed by relatively small non-diagonal coupling
$Y_{\mu \tau}^{\tilde{h}}$ ($Y_{\tau\mu}^{\tilde{h}}$). In the case of light sgoldstino or large mixing angle, sgoldstino contribution is even dominates.

\noindent
The diagrams in Fig.~\ref{sploops} containing effective vertex of
scalar and pseudoscalar sgoldstino interaction with photons are
divergent. We estimate their contributions assuming a cutoff $\Lambda$ 
for the effective theory of goldstino sector and corresponding
contribution to the Wilson coefficients in the leading order in the
$\tau$ mass reads 
\begin{eqnarray}
\label{1loopsp}
\begin{aligned}
& c_L^{sp} = \frac{M_{\gamma \gamma}}{4 m_{\tau} F^2} A_{\tau \mu} \, v \, \cos \beta \left( \log \frac{m_p^2}{m_{\tilde{s}}^2} -(1-\cos \theta) \log \frac{\Lambda^2}{m_{\tilde{s}}^2} \right)\,, \\ 
& c_R^{sp} = \frac{M_{\gamma \gamma}}{4 m_{\tau} F^2} A_{\mu \tau} \, v \, \cos \beta \left( \log \frac{m_p^2}{m_{\tilde{s}}^2} -(1-\cos \theta) \log \frac{\Lambda^2}{m_{\tilde{s}}^2} \right), \\ 
\end{aligned}
\end{eqnarray}
where $m_p$ is the mass of the pseudoscalar sgoldstino. Due to the
mixing between the scalar sgoldstino and Higgs boson, scalar and
pseudoscalar sgoldstino have different coupling constants to SM
fermions. Note, that in the absence of the mixing the result will be
finite as it was shown in Ref.~\cite{Brignole:1999gf}. Nonzero mixing
leads to a divergence in diagrams depicted in Fig.~\ref{sploops}. 
However, this divergence is only logarithmic and at the same time for
most of the models the mixing is small and the divergent part in
Eq.~(\ref{1loopsp}) is suppressed by a factor $\sim \theta^2$. For
numerical estimates we fix $m_p=200 \, \text{GeV}$ and  $\Lambda^2 =
16\pi F^2/m_{\text{sl}}^2$ which is an estimate for the scale of
perturbative violation of unitarity of the effective theory, see
the main text and detailed discussion in 
Refs.~\cite{Brignole:2000wd,Brignole:1998uu,Brignole:1996fn}.

\noindent
Finally, let us consider the contributions from 2-loop diagrams
depicted in Fig.\ref{fig:2lh}. Here we take into account only
convergent part of the 
Higgs resonance $\tilde{h}$ contribution. The divergent diagrams with
sgoldstino  are of higher order from the point of view of microscopic
theory. Moreover, we find that these diagrams are almost never
dominant; when they do dominate their contribution is
considerably smaller than the current bounds on ${\rm
  Br}(\tau\to\mu\gamma)$.  The diagrams with the internal
  $Z$-boson are suppressed by an factor of $1-4 \, s_{W}^2 \approx
0.08$ compared to diagrams with internal $\gamma$. We also neglect
them.  Finally, we are left with the contributions from upper and left
bottom diagrams on Fig.~\ref{fig:2lh}. Their contributions to Wilson
coefficients $c_{L,R}$ can be written as \cite{Harnik:2012pb}  
\begin{equation}
\label{C2loop}
c_{L,R}^{\text{2-loop}} = c_{L,R}^{t \, \gamma} + c_{L,R}^{W \, \gamma},
\end{equation}
where \cite{Harnik:2012pb}
\begin{eqnarray}
\label{C2looptg}
\begin{aligned}
& c_{L}^{t \, \gamma} = -\frac{4 \, \alpha \, G_{\scriptscriptstyle F} \,
    v}{3 m_{\tau} \pi} \, Y_{\tau \mu}^{\tilde{h}} f(z_{th}), \\ 
& c_{L}^{W \, \gamma} = \frac{\alpha G_{\scriptscriptstyle F} v}{2 m_{\tau} \pi} Y_{\tau \mu}^{\tilde{h}} \left
  [3f(z_{Wh})+5g(z_{Wh})+\frac{3}{4}g(z_{Wh})+\frac{3}{4}h(z_{Wh})+\frac{f(z_{Wh})-g(z_{Wh})}{2z_{Wh}}\right]. 
\end{aligned}
\end{eqnarray}
The loop functions $f(z),\,g(z)$ and $h(z)$ are
\begin{eqnarray}
\label{fgh}
\begin{aligned}
& f(z) = \frac{z}{2} \, \int_{0}^{1} \, dx \frac{1-2x(1-x)}{x(1-x)-z}
  \, \log \frac{x(1-x)}{z}, \\ 
& g(z) = \frac{z}{2} \, \int_{0}^{1} \, dx \frac{1}{x(1-x)-z} \, \log
  \frac{x(1-x)}{z}, \\ 
& h(z) = \frac{z}{2} \, \int_{0}^{1} \, \frac{dx}{x(1-x)-z}
  \left[1+\frac{z}{z-x(1-x)} \, \log \frac{x(1-x)}{z}\right]. \\ 
\end{aligned}
\end{eqnarray}
The same expressions for $c_{R}^{W \, \gamma}$ and $c_{R}^{t \,
  \gamma}$ can be obtained by replacement $Y_{\tau \mu}^{\tilde{h}} \to Y_{\mu \tau}^{\tilde{h}}$. 

\providecommand{\href}[2]{#2}\begingroup\raggedright
\end{document}
